**Pathways towards 30% efficient single-junction perovskite solar cells and the role of mobile ions**

*Jonas Diekmann, Pietro Caprioglio, Moritz H. Futscher, Vincent M. Le Corre, Sebastian Reichert, Frank Jaiser, Malavika Arvind, Lorena Perdigón Toro, Emilio Gutierrez-Partida, Francisco Peña-Camargo, Carsten Deibel, Bruno Ehrler, Thomas Unold, Thomas Kirchartz, Dieter Neher, Martin Stolterfoht[*]*

Jonas Diekmann, Pietro Caprioglio, Dr. Vincent M. Le Corre, Dr. Frank Jaiser, Malavika Arvind, Lorena Perdigón Toro, Emilio Gutierrez-Partida, Francisco Peña-Camargo, Prof. Dr. Dieter Neher, Dr. Martin Stolterfoht

Institute of Physics and Astronomy, University of Potsdam, Karl-Liebknecht-Str. 24-25, D-14476 Potsdam-Golm, Germany
E-mail: stolterf@uni-potsdam.de

Dr. Moritz H. Futscher, Prof. Dr. Bruno Ehrler
AMOLF, Center for Nanophotonics, Science Park 104, 1098 XG Amsterdam, The Netherlands

Sebastian Reichert, Prof. Dr. Carsten Deibel
Institut für Physik, Technische Universität Chemnitz, 09126 Chemnitz, Germany

Dr. Thomas Unold
Department of Structure and Dynamics of Energy Materials, Helmholtz-Zentrum-Berlin, Hahn-Meitner-Platz 1, D-14109 Berlin, Germany

Prof. Dr. Thomas Kirchartz
IEK5-Photovoltaik, Forschungszentrum Jülich GmbH, 52425 Jülich, Germany

Perovskite semiconductors have demonstrated outstanding external luminescence quantum yields, enabling high power conversion efficiencies (PCE). However, the precise conditions to advance to an efficiency regime above monocrystalline silicon cells are not well understood. Here, we establish a simulation model that well describes efficient *p-i-n* type perovskite solar cells and a range of different experiments. We then study important device and material parameters and we find that an efficiency regime of 30% can be unlocked by optimizing the built-in potential across the perovskite layer by using either highly doped ($10^{19}$ cm$^{-3}$), thick transport layers (TLs) or ultrathin undoped TLs, e.g. self-assembled monolayers. Importantly, we only consider parameters that have been already demonstrated in recent literature, that is a bulk lifetime of 10 μs, interfacial recombination velocities of 10 cm/s, a perovskite bandgap ($E_{\text{gap}}$) of 1.5 eV and an EQE of 95%. A maximum efficiency of 31% is predicted for a bandgap of 1.4 eV. Finally, we demonstrate that the relatively high mobile ion density does not represent a significant barrier to reach this efficiency regime. Thus, the results of this paper promise continuous PCE improvements until perovskites may become the most efficient single-junction solar cell technology in the near future.

**1. Introduction**

Over the last 10 years perovskite solar cells have triggered an enormous research interest and with power conversion efficiencies of 25.5%[1] they are close to the efficiencies of monocrystalline silicon solar cells (26.7%).[2,3] As such, perovskites provide an exciting opportunity to approach the thermodynamic efficiency limit of single-junction perovskite solar cells (33.7%[4,5]) due to their exceptional optical and material properties. An optimal semiconductor for solar cell applications must basically fulfil three requirements, that is (i) a high absorption coefficient, in particular strong absorption at the band edge, (ii) long charge carrier diffusion lengths to efficiently collect charges in several hundred nm thick films and (iii) a high photoluminescence quantum yield (PLQY).[6] Despite the impressive efficiencies achieved with perovskite solar cells, there are many other key considerations that need to be addressed prior to a successful commercialization of the technology, such as the environmental stability,[7] the toxicity of lead,[8] and the scalability of the fabrication processes.[9] Notwithstanding these points, in order to improve the efficiencies of state-of-the-art perovskite cells further it is imperative to supress all non-radiative recombination to an absolute minimum. This is particularly the case for the perovskite/transport layer (TL) interfaces but also for the perovskite absorber itself. This task requires a deep understanding of defect chemistry and device physics. Very encouraging in this regard are recent studies which highlighted exceptional external PL efficiencies (PLQY) in several perovskite compositions.[10–15] For example Braly et al.[11] demonstrated an external fluorescence yield of 20% with a charge carrier bulk lifetime ($\tau_\text{bulk}$) of approximately 8.8 µs, even with the most standard methyl ammonium lead iodide (MAPI) perovskite absorber which was passivated with trioctylphosphine oxide (TOPO).[16,17] Even higher PLQYs were demonstrated on a neat, potassium-passivated mixed perovskite film, however, no lifetimes were specified.[10] It is also clear, that when it comes to complete cells, a key consideration is the non-radiative loss at the interfaces. A suitable approach to quantify the non-radiative recombination loss at the interface is based on absolute photoluminescence.[18–20] In fact, many recent record cells now commonly report luminescence quantum efficiencies of several percent which underline the exceptional opto-electronic quality of the whole device stack.[13] Important parameters that define the non-radiative interfacial recombination loss are the interface recombination velocity ($S$) and the energy offsets between the TLs and the perovskite.[21,22] Both parameters depend on the underlying substrate and the specific chemical environment at the interface.[23] In very high efficiency systems, $S$ can be reduced to be on the order of 10 cm/s[12,22,24] although this parameter is not routinely quantified in the literature and further research must be done to clearly disentangle the interfacial charge transfer and recombination based on kinetic measurements, such as transient PL.[20,22,24–27]

A key consideration to improve the efficiency of single junction cells is to lower the bandgap ($E_{gap}$) to the optimum value in the Shockley-Queisser model (e.g. ~1.34 eV).[4,5] Commonly used perovskites such as methylammonium lead iodide (MAPI)[28,29] and formamidinium lead iodide (FAPI)[30–32] do allow some flexibility in modifying the bandgap within a range of 1.6 eV to approximately 1.5 eV,[33] while $CsPbBr_3$ allows to increase the bandgap further to 2.4 eV[34] which is an important consideration for tandem solar cell applications.[23] Notably, both MAPI and FAPI perovskite cells allow achieving PCEs above 20%.[20,29,35] Moreover, the incorporation of small amounts of Sn[36–38] allows reducing the bandgap to the optimum bandgap in the S.Q. model (1.1-1.4 eV) and PCEs above 21% have been reported for cells based on a 1.22 eV perovskite bandgap in 2019.[39] Unfortunately, tin perovskite brings other problems due to the high probability of oxidation from $Sn^{2+}$ to $Sn^{4+}$.[38] Other key parameters include the donor and acceptor density ($n_{dop}$) in the TLs (i.e. doping),[40–42] the charge carrier mobilities ($\mu$) in all layers,[43] as well as the kinetics and density of mobile ions ($n_{ions}$), which are expected to screen and re-distribute the internal field thereby causing recombination losses and hysteresis effects.[35,44–47] Other important considerations include the built-in potential ($V_{BI}$) across the device and the energy-level alignment between the perovskite and the charge TLs for majority carriers ($\Delta E_{maj}$).[48–50] It is fair to say that until today the impact of most of these parameters remains heavily debated in the community and there exist many conflicting findings in the literature with regard to these parameters. For example, so far, no consensus has been reached on the density of mobile ions with reported values typically varying from $10^{15} - 10^{19}$ cm$^{-3}$ for MAPI.[46,51–53] Moreover, some literature reports suggest that the energy-level alignment does not play an important role for the device performance,[54] which is in contrast to other studies where the $V_{OC}$ correlated to $\Delta E_{maj}$.[25,55–57] Also, the distribution of the built-in potential ($V_{BI}$) across the device remains in general poorly understood despite significant progress in this regard.[35,58–60] Other examples include the chemical doping with several reports linking doping to enhanced losses in $V_{OC}$ or quasi-Fermi level splitting (QFLS),[61–63] although in many inorganic solar cell technologies doping is a key factor in preventing minority carrier recombination at the metal contacts.[3,64]

In this work, we aim to understand the most promising and simple optimization strategies in order to allow perovskites to reach and surpass the efficiency of monocrystalline silicon and even GaAs (29.1%).[1] To this end we have optimized previously validated device simulations based on the well-established drift-diffusion simulator SCAPS,[65,66] by fitting the simulations to a number of different experimental results. Furthermore, the numerical simulations were extended to incorporate mobile ions based on experimentally measured diffusion constants and ion densities using the software IonMonger.[67,68] Based on a set of standard simulation parameters that describe a typical *p-i-n* type cell (with a PCE of 19.2%), we have varied several important parameters to understand their relative importance within the limits of the simulation model, such as $n_{dop}, n_{ions}, \mu, V_{BI}, E_{maj}, E_{gap}, S, \tau_{bulk}$ and the device absorption. Starting from the standard simulation, we find that increasing the carrier mobilities (in all layers) by a factor of 10 may only lead to small efficiency improvements (~1% absolute) by allowing fill factors of up to 85%. This is in contrast to chemical doping of the transport layers ($n_{dop} = 10^{19}$ cm$^{-3}$) which allows exceeding FFs of 85% and to reach $V_{OC}$s close to the radiative limits with PCEs of 24% for the standard cell configuration. We explain the potentially huge benefit of doped TLs through an effective increase of the $V_{BI}$ across the absorber layer which repels minority carriers from the interfaces, thereby reducing interfacial recombination losses. Related to this, we show that a high dielectric constant in the neat material would lead to a decrease in fill factor due to a decrease in the $V_{BI}$ across the absorber layer. These considerations imply the importance of maximizing the $V_{BI}$ across the perovskite and also imply the benefit of a small ion density to prevent a screening of the internal field ($n_{ion}$ << electrode charge/volume ~$10^{16}$ cm$^{-3}$). Nevertheless, the experimentally measured mobile ion density greatly exceeds this critical value ($\geq 6 \times 10^{16}$ cm$^{-3}$) and we discuss the implication of this on the device performance. Finally, we demonstrate that 30% efficient perovskite solar cells are within reach by only using parameters that have already been demonstrated in recent literature, i.e. a bandgap of approximately 1.5 eV, a bulk carrier lifetime of 10 µs, an interfacial recombination velocity of 10 cm/s and an EQE of 95% throughout the spectrum. The key to unlock this efficiency regime is to maximize the device built-in field by either increasing the doping concentration in the TLs to at least $10^{19}$ cm$^{-3}$, or using ultrathin (≈10nm), undoped TLs, if a strong built-in field can be created by other means. Finally, we show that although the mobile ion density is very significant in our triple cation perovskite devices ($\geq 6 \times 10^{16}$ cm$^{-3}$), it will most likely not represent a significant barrier for reaching such near-ideal solar cell performances as the relative impact of the mobile ions decreases with reduced interfacial and bulk recombination.

## 2. Results

In the first part of the study, we optimized our SCAPS[65,66] device simulation parameters by fitting several experimental results. First, standard simulation parameters were established to describe our reference cells with the following architecture: ITO/PTAA/PFN-Br/perovskite/C$_{60}$/BCP/Cu.[20,69] These cells exhibit on average a $V_{OC}$ of approximately 1.13 V, a fill factor of 79% and a short-circuit current density of 21.5 mA cm$^{-2}$ with an average external quantum efficiency (EQE) of close to 88% between 400 nm to 750 nm (**Supplementary Figure S1**) which results in a PCE of ~19.2%.[20] We note these *p-i-n* type cells initially exhibit essentially no hysteresis at different scan speeds that are accessible with our *JV*-measurement setup (scan rate <150 mV/s, **Supplementary Figure S2**), which, however, doesn't imply the absence of a hysteresis at even faster scan speeds. Using TRPL measurements, we obtained an interface recombination velocity of $S = 2000$ cm/s at the perovskite/ETL interface and $S = 100$ cm/s at the HTL/perovskite interface, as well as a bulk lifetime $\tau_{bulk}$ of 500 ns (**Supplementary Figure S3**).[20] Moreover, we note that the perovskite layer is effectively undoped ($\approx 6\times 10^{11}$ cm$^{-3}$) as evidenced by lateral conductivity measurements (**Supplementary Figure S4**). A fit of the standard simulations to experimental *JV*-curves is shown in **Figure 1a** while the simulation parameters are shown in **Supplementary Table S1.** The set of basic equations solved by SCAPS is presented in **Supporting Note 1.** We note, that we used a bandgap-dependent bimolecular recombination coefficient ($k_2$) as detailed in **Supporting Note 2.** Knowing the radiative recombination current density in the dark and the effective density of states $N_C = N_V = 2.2\times10^{18}$ cm$^{-3}$,[24] $k_2$ can be readily obtained (e.g. for a bandgap of 1.63 eV, we obtain a $k_2 = 3\times10^{-11}$ cm$^3$/s).[70,71] We note, that this represents the external $k_2$ which is impacted by photon recycling inside the bulk.[6,72–75] Therefore, by using the external $k_2$, we effectively consider the effect of photon recycling in our simulations. Importantly, using the bandgap-dependent $k_2$ and no parasitic losses, we can well reproduce the Shockley-Queisser efficiency vs. bandgap (**Supplementary Figure S5**) with a maximum PCE of 33.77% at a bandgap of 1.36 eV.[5] Note, the precise value of $k_2$ has no impact under a 1 sun illumination for cells which are limited by SRH recombination in the bulk or at the interfaces (such as the standard cell). Moreover, it is interesting to note that reported Auger recombination coefficients ($10^{-28}$ cm$^{-6}$/s)[70] reduce the maximum obtainable PCE in our simulations by only 0.03% and are therefore not a limiting factor. The simulated parallel recombination currents in the perovskite layer, interface and/or metal contacts are shown in **Figure 1b** which accurately correspond to the experimentally obtained recombination currents (**Figure 1c**). The latter were obtained by measuring the QFLS of individual perovskite/transport layer films of the cell as reported in a previous publication.[48] Notably, at $V_{OC}$, interfacial recombination outweighs the recombination in the neat material by more than 1 order of magnitude. Moreover, the simulations reproduce the ideality factor of the standard cells of approximately 1.35 across a broad range of light intensities (**Figure 1d**).[76] Based on these settings, several recent experimental results were fitted by changing only one parameter depending on the particular experiment. These experiments include cells with reduced interface recombination at the perovskite/ETL interface upon adding LiF (**Figure 1e**),[20] cells with different hole transport layers resulting in different $V_{OC}$s (**Figure 1f**),[48] as well as cells with different PTAA layer thicknesses resulting in different fill factors (**Figure 1g**).[43,77] Overall, Figure 1 demonstrates that the SCAPS simulations can well reproduce these experimental results while also produce the quantified recombination currents in the bulk and the interfaces by taking into account the measured interface and bulk lifetimes.

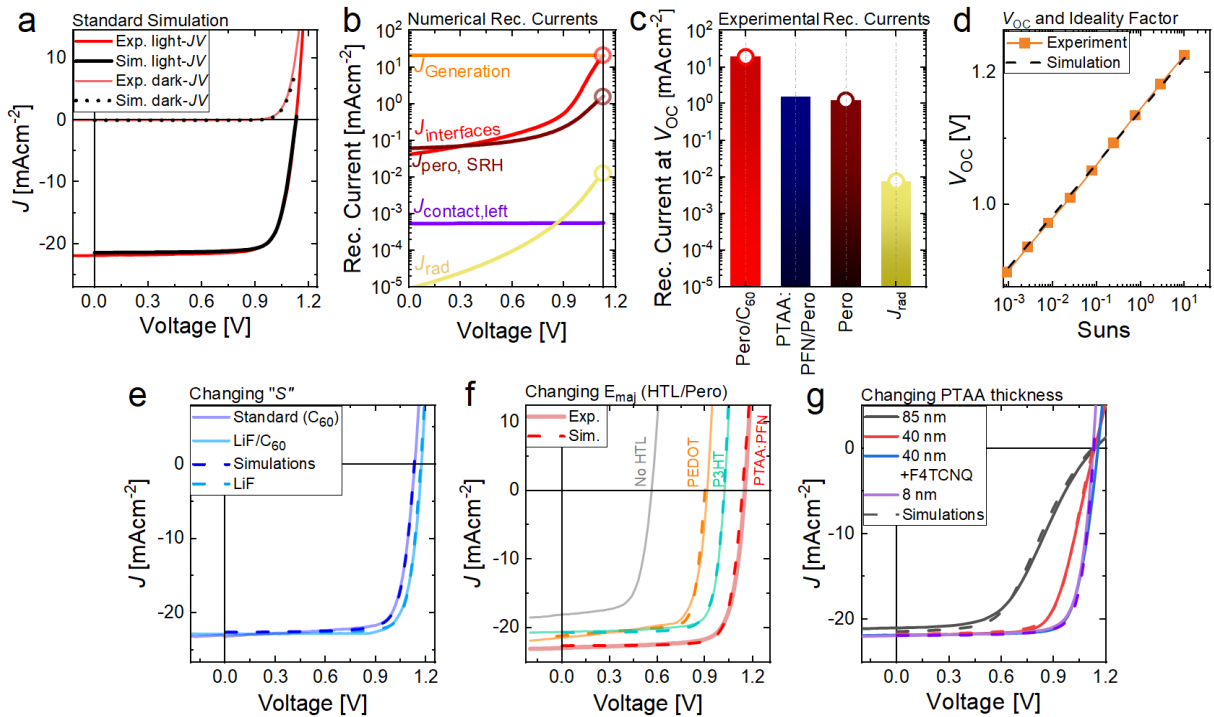

Figure 1. (a) Experimental and numerically simulated light and dark JV curves of p-i-n type cells based on PTAA/PFN-Br/triple cation perovskite/$C_{60}$. The interface recombination velocities and bulk lifetime were obtained from transient PL (TRPL) measurements.[20] (b) Simulated voltage-dependent radiative and non-radiative recombination currents in the neat perovskite and at both interfaces compared to the generation current. Interfacial recombination reduces the possible open-circuit voltage which is roughly 10× larger than Shockley-Read-Hall recombination in the neat material. (c) This is consistent with experimentally measured recombination currents under open-circuit conditions using steady-state PL measurements.[48] (d) The simulation parameter set also reproduces the ideality factor ($n_{ID} \sim 1.35$) of the standard cells.[76] The standard simulation settings allow to fit several experimental results by changing a single parameter depending on the particular experiment. (e) Inserting LiF between the perovskite and $C_{60}$ leads to concurrent increase of the transient PL (TRPL) lifetime, device $V_{OC}$

and PCEs above 21%.[20] The TRPL results suggest that S decreases 1 order of magnitude upon application of LiF which allows to reproduce the experimentally observed $V_{OC}$ gain of 35 mV. **(f)** Cells with different HTLs and varying majority carrier band offsets as obtained from UPS measurements.[48] Implementing energy offsets at the p-interface for the cells with a PEDOT:PSS and P3HT hole transport layer allows to reproduce the experimental JV-curves. **(g)** Cells with changing PTAA thickness. The mobility in the PTAA layer has a significant impact on the device fill factor consistent with previous work.[77] The fitting allowed to finetune the perovskite and the PTAA layer mobility ($\mu_{pero}$=1 cm$^2$/V/s, $\mu_{PTAA}$=1.5×10$^{-4}$ cm$^2$/V/s, while $\mu_{C60}$=1.0×10$^{-2}$ cm$^2$/V/s).

## 2.1 Mobile ions

Despite the good fits shown in **Figure 1**, it is important to note that the SCAPS simulations above were performed assuming a negligible mobile ion density.[43–46] To address the role of mobile ions on our simulation results, we performed additional experiments and simulations using the open-source simulation software IonMonger.[59,60] In a first step, we confirmed the comparability of IonMonger and SCAPS (without mobile ions) by comparing the key-parameters such as $S$ and the majority carrier offset $E_{maj}$ (**Supplementary Figure S6**) which provided very similar results. To implement the mobile ion density in IonMonger, we first quantified the ion diffusion constant and activation energy ($E_A$) using transient capacitance measurements following a recently introduced approach (**Supplementary Figure S7**).[46,51,78] We note that we observed in fact the features of 2 mobile ion species, however, for simplicity we only considered the faster component in the simulations ($D_{ion}$ = 6×10$^{-10}$ cm$^2$/s and $E_A$ = 120 meV). Moreover, in order to estimate the ion density in our cells, we used bias-assisted charge extraction (BACE) experiments (**Supplementary Figure S8**). The BACE measurements indicate an ion density of 6x10$^{16}$ cm$^{-3}$ at a collection voltage of -1.6 V, which is consistent with Bertulozzi et al. who obtained a value of 7x10$^{16}$ cm$^{-3}$ for double cation/halide perovskite compositions (e.g. Cs$_{0.17}$FA$_{0.83}$Pb(I$_{0.83}$Br$_{0.17}$)$_3$). Unfortunately, as larger collection voltages would break the devices, we can presently not exclude that the ion density is larger than the obtained value. **Figure 2a** and **b** visually depict the obtained $V_{BI}$ distribution in steady-state from IonMonger upon implementing a mobile ions density of 6x10$^{16}$ cm$^{-3}$ (assumed to be halide vacancies), in comparison to the idealized homogenous distribution obtained in SCAPS. **Figure 2c-e** demonstrates a good agreement between both simulation programs in terms of the *JV*-scan, bulk and interface recombination currents as well as ideality factor. We note that we used similar simulation parameters for both programs, however, in order to compensate the detrimental effect of the significant ion density in IonMonger, we used a slightly higher built-in voltage (+100 mV). Overall, this comparison shows that both programs provide very similar results and although the mobile ion density significantly redistributes the $V_{BI}$, its effect on the device performance is rather small (i.e. ~1.5% in PCE), which is obtained by comparing the PCE in case with and without mobile ions in IonMonger.

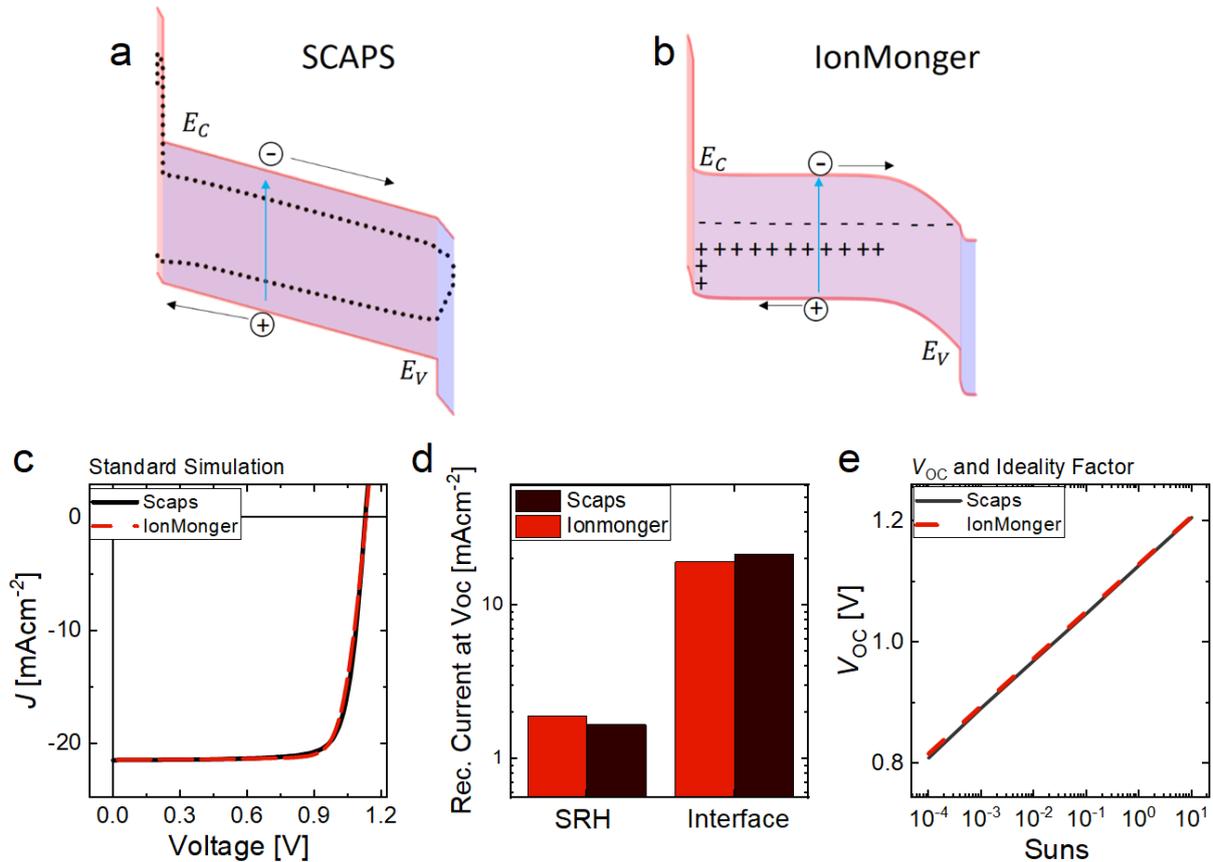

**Figure 2.** **(a)** and **(b).** The band-diagrams and resulting built-in potential obtained in SCAPS and IonMonger at short-circuit conditions. In case of Ionmonger, the signs in the active layer denote the distribution of mobile halide vacancies ("+") and immobile remaining ions ("–").

**(c)**, **(d)** and **(e)** show a direct comparison of the JV, interfacial and bulk recombination currents and the ideality factor for both programs.

Starting from this simulation set, we aim to find the most promising ways to improve the PCE of perovskite solar cells. A large set of simulations was performed to check most parameters accessible in SCAPS and IonMonger. We note, that in the following, we will show the results as obtained from SCAPS, while IonMonger simulations are shown again towards the end of the paper. We also emphasize that we have cross-checked each simulation with IonMonger which led to very similar results and identical trends.

### 2.2 Mobility and TL Doping

A generic approach to optimize the performance of a solar cell is to maximize the carrier mobility to improve the charge-extraction efficiency. To this end, we first varied the majority carrier mobility in the HTL and ETL simultaneously as well as both carrier types in the perovskite layer (**Supplementary Figure S9**). Only relatively small PCE improvements (∼1.5%) are possible by increasing the mobility in all layers (**Figure 3a**). We note that this improvement comes likely from the hole mobility in the HTL because the PTAA layer mobility ($\mu_{\text{PTAA}}=1.5\times10^{-4}$ cm$^2$/V/s) with respect to the thickness of the layer ($d_{\text{HTL}}=10$ nm) is significantly lower than the C$_{60}$ and perovskite mobility ($\mu_{\text{C60}}=1.0\times10^{-2}$ cm$^2$/V/s, $\mu_{\text{pero}}=30$ cm$^2$/V/s) vs. their thicknesses (30 nm and 500 nm, respectively). Therefore, PTAA sets the transport limitation despite the thin layer.

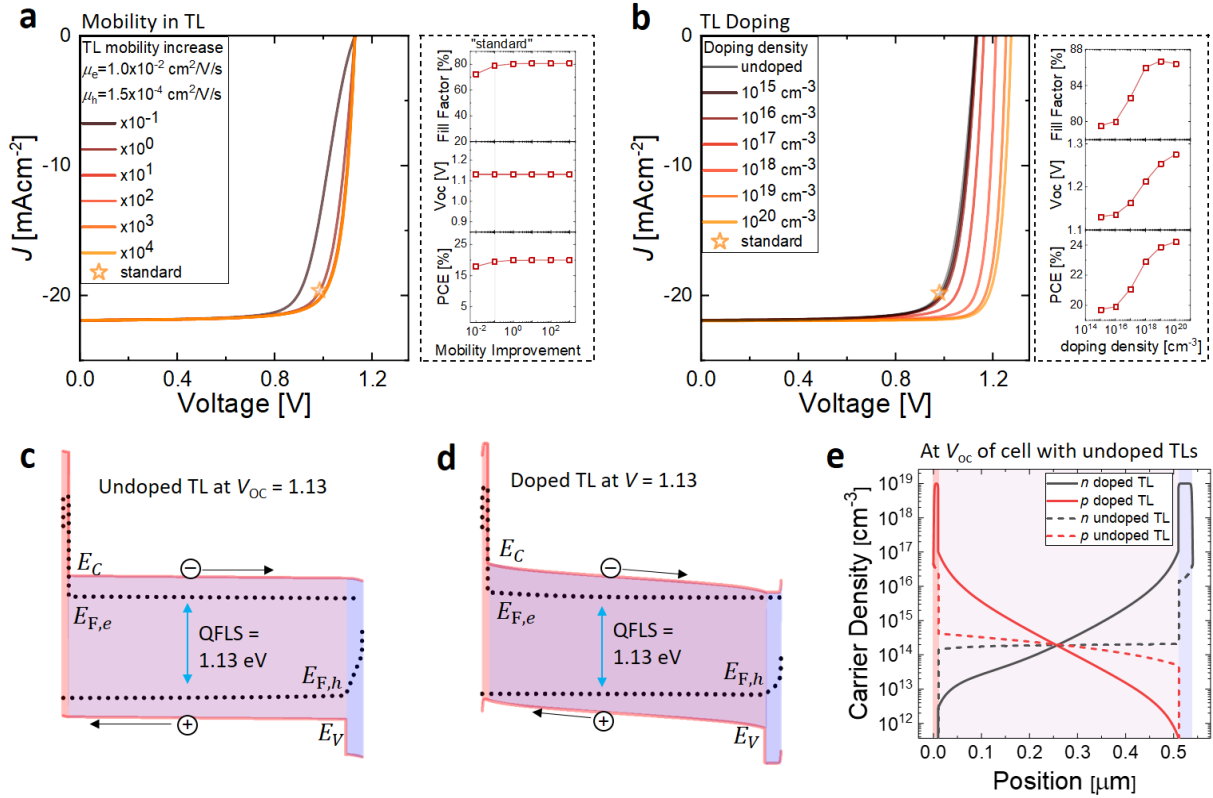

**Figure 3. (a)** Numerically simulated current density vs. voltage characteristics of p-i-n type perovskite solar cells by simultaneously increasing the e$^-$ and h$^+$ mobility ($\mu_{e^-}$, $\mu_{h^+}$) in the transport layers (TLs) by the specified factor in the legend, which allows relatively small efficiency improvements (< 1%). **(b)** By contrast, increasing the donor/acceptor doping density in the TLs (and keeping the mobility constant) can significantly increase the fill factor (FF) and $V_{OC}$ of the cell. **(c)** and **(d)** show the resulting band diagrams of a cell with undoped and doped TLs (10$^{19}$ cm$^{-3}$), respectively. Doping increases the effective driving field across the active layer which repels minority carriers from the interfaces. The resulting carrier densities at the $V_{OC}$ of the undoped, reference cell (1.13 V) are shown in panel **(e)**.

The situation is, however, markedly different upon chemical doping of the TLs. **Figure 3b** shows that both the FF and the $V_{OC}$ improve significantly by increasing the donor/acceptor concentration in the TLs (from 10$^{15}$-10$^{20}$ cm$^{-3}$). This results in PCEs above 24% for the standard cell. We attribute this result to the effective increase of the built-in-field across the perovskite layer which greatly reduces the minority-carrier concentration at the interfaces and thus interfacial recombination. This is because of two effects: 1) The potential drop across heavily doped TLs becomes negligible and 2) the energy offset of the doped TLs (difference between the ionization potential of the HTL and electron affinity of the ETL) maximizes the voltage drop across the perovskite (here 1.6 V). This is illustrated by the band diagrams in **Figure 3c,d** for an undoped (reference) cell at its $V_{OC}$ as well as the same cell with doped TLs ($n_{dop}=10^{19}$ cm$^{-3}$). The latter still exhibits a considerable internal field at the same applied voltage. **Figure 3e** shows that these concomitant effects lead to a significant reduction of the minority-carrier density at the interfaces which supresses the dominating interfacial recombination (see **Supplementary Figure S10a**) and increases the $V_{OC}$ and the FF. Interestingly, the higher carrier density in the doped TL does not speed up the SRH recombination, because the recombination rate is largely determined by the minority-carrier density at the interfaces. In general, the increase in the TL conductivity also reduces charge transport losses, although for our reference cell this effect is small due to the thin TLs and little transport losses (**Figure 3a**). The relative importance of these effects is further discussed in **Supplementary Figure S10b.** It is also interesting to note that both TLs need to be doped simultaneously as otherwise (if only one TL is doped) the other interface becomes the limiting component of the cell. Therefore, doping of the $C_{60}$ layer allows considerable improvements until the PTAA/perovskite interface and the absorber layer itself are limiting the performance (**Supplementary Figure S11**). In contrast, almost no performance gains are achieved by doping the HTL only, as in this case the perovskite/$C_{60}$ interface remains the limiting factor for the FF and $V_{OC}$. Lastly, we emphasise that identical performance improvements can be achieved by using only a thin doped interlayer instead of a doped TL as shown in **Supplementary Figure S12**, which increases the experimental possibilities to reduce the critical interfacial recombination at the top-interface through implementation of a doped interlayer.

**2.3 Device built-in potential & energy-level alignment**

In a next step, we aimed to understand the impact of the *built-in electrostatic potential difference* ($V_{BI}$) and the *energy-level alignment* ($E_{maj}$) on cells with doped and undoped TLs.

Both parameters have been shown to be critical for the device performance.[25,48–50,55,79] **Figure 4a** shows that a high built-in voltage above 1.2 V across the whole device is required in the undoped cell with a band gap of 1.63 eV in order to efficiently extract the carriers and minimize fill factor losses and to prevent the formation of a reverse field in the device.[79] A similar picture appears to be the case for the cell with doped TLs ($n_{dop}=10^{19}$ cm$^{-3}$), although the cell is more tolerant to a lower $V_{BI}$ with >0.8 V being sufficient to avoid considerable PCE and FF losses. This is also shown in **Supplementary Figure S13** for cells with and without doped TLs for different $V_{BI}$ and $E_{maj}$ values. In fact, we believe that in reality no metal work function mismatch may be required in case of strongly doped TLs, because carriers could tunnel through the TL to the electrode which is challenging to implement numerically. Therefore, the observed PCE drop in **Figure 4a** at $V_{BI}$s below 0.8 V may be incorrect. For the reference cell with undoped TLs, we find that $V_{BI}$s below 1.0 V would not allow reproducing our high FFs of close to 80% even if the carrier mobilities are significantly increased (**Supplementary Figure S14**). Note, that we assumed $V_{BI} = 1.2$ V in the standard settings which is consistent with Mott-Schottky analysis in our cells. However, we acknowledge that we do not precisely know its origin considering the almost equal work function of ITO and Cu.[48] A large $V_{BI}$ may indicate a considerable modification of the metal work functions in the presence of thin organic layers and/or the perovskite layer, or that the built-in potential is a result of remotely doped TL in our cell.[80,81]

While a built-in potential across the complete device (in the dark) is indeed necessary to achieve a well-performing device, it is an interesting question whether it is possible that the $V_{BI}$ drops only across the transport layers but not across the perovskite layer, as it is sometimes assumed in literature. To investigate this, we artificially increase the dielectric constant in the perovskite layer which screens the field in the perovskite, thereby redistributing the relative potential drop across the perovskite layer and the TLs. **Supplementary Figure S15** shows that when the mobilities in the TLs are kept constant, the efficiency decreases the larger the relative drop of the built-in voltage across the transport layers which reduces the voltage across the perovskite. This efficiency loss occurs even if the diffusion length in the perovskite layer ($L_d$) exceeds the film thickness (*d*) as observed experimentally.[82–84] The reason is that the charge extraction may not be limited by the perovskite layer but rather by the TLs, and also because interfacial recombination (which also impacts the FF) would be considerably enhanced if the $V_{BI}$ was lower across the perovskite layer. Nevertheless, this does not mean that the potential cannot be partially flat across the perovskite active layer. For example, as shown above at **Figure 2b**, in case of a significant mobile ion density, large field-free regions do exist in the perovskite layer, however, there is also a significant field at perovskite surfaces that repels minority carriers from the interface. This scenario considerably lowers the interfacial recombination, which means that the majority of the perovskite bulk could be field-free provided that the charge carrier diffusion length in the perovskite is sufficient.

As to the impact of the energy level alignment, **Figure 4b** shows that the energy levels of the transport layers need to be matched with respect to the energy levels of the perovskite layer and any downhill energetic offset for electrons (uphill for holes) would cause substantial $V_{OC}$ and PCEs losses regardless whether the TLs are doped or not. Considering our previous study where we observed a match between the internal QFLS and the device $V_{OC}$,[47] we expect that the alignment in our standard cells is well optimized. We note that any energetic offset will cause an equal loss in device $V_{OC}$ as long as the interface between the perovskite and the misaligned TL is limiting the performance of the cell (and not another interface or the bulk).

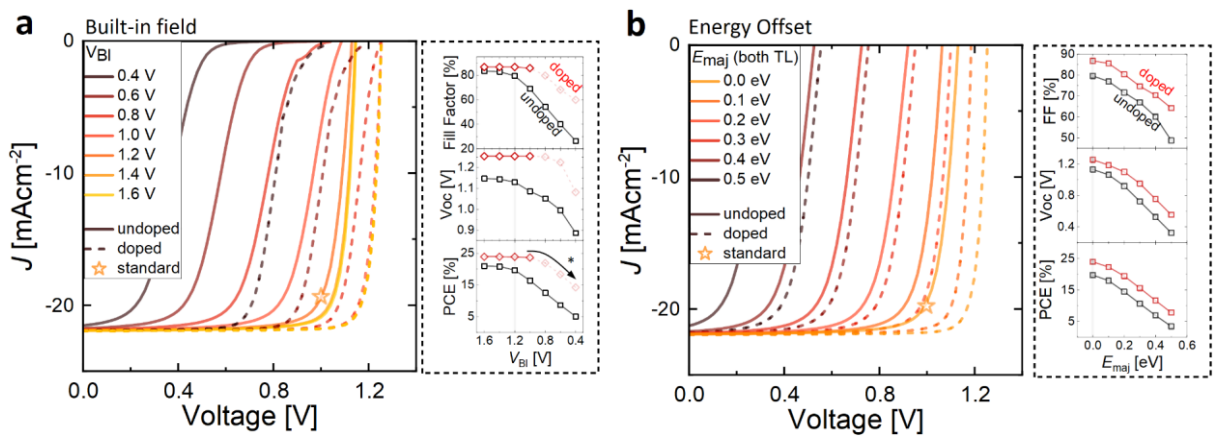

**Figure 4. (a)** The impact of the device built-in potential ($V_{BI}$, here defined as the work function difference of the contact metals) and **(b)** the energy alignment between the perovskite and the transport layers (TLs) on the JV-characteristics and performance parameters of cells with doped and undoped TLs. In panel (b), the $V_{BI}$ was kept constant (1.2 V). In the case of intrinsic TLs, increasing the $V_{BI}$ from 1.2 V (standard cell) to 1.6 V allows to improve the FF to 83.5%. Doped TLs are more tolerant with respect to the $V_{BI}$. Considerable losses appear below 1.0 V (marked with *), however, we attribute this to a limitation of the simulation model as discussed in the main text. In contrast, the energy level mismatch between the perovskite and the transport layers has a large impact on the device $V_{OC}$ regardless of the doping of the TLs.

## 2.4 Suppressed Recombination

Having analysed the impact of the charge-carrier mobility, doping, mobile ions and the energetics on the device performance, we now focus our attention on supressing the defect recombination in the bulk of the material and at the interfaces.[85] In the following we demonstrate the impact of $S$ and $\tau_{bulk}$ as a function of the perovskite bandgap. **Figure 5a** shows the PCE of the standard cell as a function of the perovskite bandgap (red line). Interestingly, lowering the bandgap (from currently 1.63 eV to the optimum of ~1.34 eV[5]) does not lead to significant performance gains (from 19.2% to 20.4%) due to the limitations imposed by the interfaces and the recombination in the bulk. We note that we kept the injection barriers between the contact metals and the TLs at 0.2 eV (as used in the standard simulations) and the energy levels of the transport layers well aligned, as otherwise the PCE drops rapidly for higher and lower bandgaps (see **Supplementary Figure S16**). This also indicates difficulties in maintaining the perovskite performance when increasing the perovskite bandgap when the energy levels of the TLs do not move in accordance with the energetics of the perovskite absorber (**Supplementary Figure S17**). Although this may actually not be the limiting mechanism (considering a possible phase segregation, see ref.[34]), it could be relevant when aiming for high performance Si/perovskite tandem cells where a perovskite bandgap of 1.7 eV would be better than a typical gap of 1.6 eV.[86] **Figure 5a** also shows that switching off bulk recombination (blue) does not improve the PCE across the bandgap as it is the interface recombination that governs the PCE. Improvements up to a PCE of 23.8% are, however, possible if interface recombination is switched off completely (green); and close to 27% when both $S$ and $\tau_{\text{bulk}}$ are switched off simultaneously (orange, "bimolecular only"). Considering now optimized, yet already demonstrated recombination parameters, e.g. $S$ = 10 cm/s[12,22,24] (at both interfaces) and $\tau_{\text{bulk}}$ =10 µs (light blue)[16] allows reaching efficiencies of 25.3% which is above the situation without interfacial recombination. Interestingly, the optimum bandgap in case of the optimized recombination parameters is considerably different (~1.36 – 1.53 eV) than expected from the S.Q. model (1.12 – 1.4 eV). This is indicated by the dashed circles in **Figure 5a**.

In the following we tried to maximize the PCE using undoped TLs. First, we note that all simulations so far were performed by using an absorption model for a direct semiconductor which matches the experimentally measured absorption coefficient, $\alpha$ and reproduces the short-circuit current in our standard cells with an average EQE slightly below 88% (**Supplementary Figure S1**). By further increasing the light-in-coupling and thus the EQE to 95% above the bandgap while simultaneously enhancing the carrier mobilities in all layers by a factor of 10 and also reducing the injection barriers from the metals to the TLs from 0.2 eV to 0.1 eV (i.e. increasing the $V_{\text{BI}}$), we obtain a maximum efficiency of 29.4% at a bandgap of 1.36 – 1.4 eV. To check whether such high EQEs are reasonable in our devices, we also performed transfer-matrix simulations[87,88] (see **Supplementary Figure S18** and corresponding discussion) which predict an average above-gap EQE slightly below 95% resulting in a $J_{SC}$ that is ~1 mAcm$^{-2}$ lower than the current obtained with SCAPS for the best case, marked with an open star in **Figure 5a**. Nevertheless, we note that EQE spectra with an average EQE of approximately 95% above the bandgap have been already demonstrated in efficient *n-i-p*-type devices with certified short-circuit currents and PCEs (23.48%).[89] This is clearly a key aspect to reach the 30% PCE milestone.

In the next step, starting from the simulation with optimized recombination parameters (light blue, max PCE = 25.3%), we further checked the impact of TL doping (using again 10$^{19}$ cm$^{-3}$ for both TLs). As shown in **Figure 5b**, this further enhanced the PCE to 28.7% (dashed dotted light blue). As a comparison the red lines show the corresponding PCE improvement vs. $E_{gap}$ upon TL doping in the standard cells. Finally increasing the EQE to 95%, results in a maximum PCE of 31% at a bandgap of 1.4 eV (purple line in **Figure 5b**). We note that for the same simulation parameters, the PCE stays above 30% for bandgaps of up to 1.5 eV, yet drops to 27.7% at a bandgap of 1.6 eV. We also checked under which conditions, in particular, for which energetic offsets and interface recombination velocities such high power conversion efficiencies may be maintained. **Figure 5c**, **d** and **e** shows the PCE as a function of *S* and $E_{maj}$ at both interfaces for our standard cells, as well as the optimized cells with undoped and doped TLs, respectively. Interestingly, **Figure 5e** shows that PCEs of ~30% may be maintained in case of doped TLs even with interface recombination velocities of 1000 cm/s if the majority carrier band offsets are small ($E_{maj}$ between -0.1 eV and 0 eV). However, higher values and positive energy offsets lead to rapid efficiency losses. We note that an analogous plot of PCE vs. *S* and the bulk lifetime is shown in **Supplementary Figure S19.**

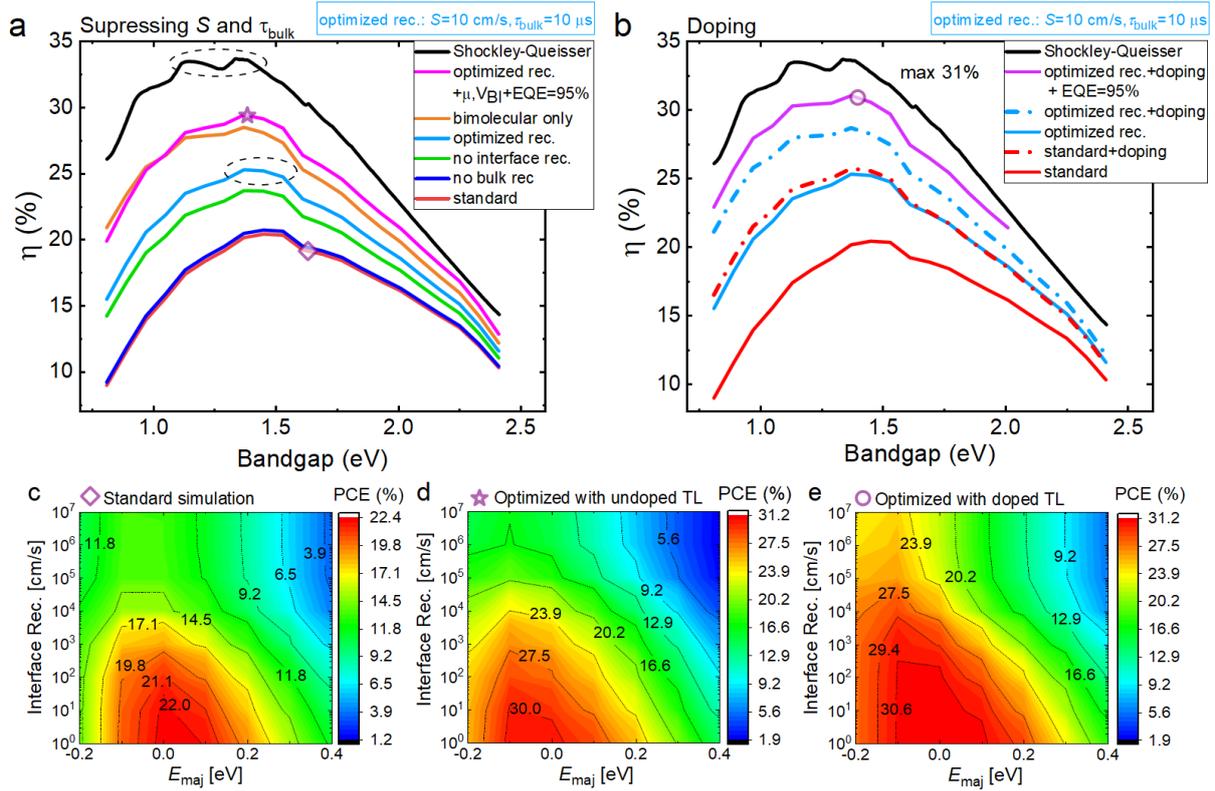

**Figure 5. (a)** Numerically simulated power conversion efficiency (PCE) versus perovskite bandgap ($E_{gap}$) considering aligned perovskite/transport layer energy levels for different scenarios. Starting from the standard settings plotted in red (S = 2000 cm/s at the perovskite/ETL interface and bulk lifetime of 500 ns); (blue) in absence of bulk recombination; (green) in absence of interfacial recombination; (orange) in absence of interface and bulk recombination (bimolecular recombination only); (light blue) using optimized, yet plausible recombination parameters (S=10 cm/s at both interfaces and a bulk-lifetime, $\tau_{bulk}$, of 10 µs). The magenta curve depicts an optimized cell with undoped TLs with optimized bulk and interface recombination, increased EQE (95% above gap), increased mobilities in all TLs by a factor of 10 and reduced injection barriers from the metal to the TLs (from 0.2 eV to 0.1 eV) resulting in a PCE of 29.4% at an optimum gap. **(b)** shows the impact of TL doping ($10^{19}$ cm$^{-3}$) for the standard cell in red (solid line without doping, dashed dotted line with doping), the cell with optimized recombination parameters in light blue (solid line without doping, dashed dotted line with doping), as well as for the cell with optimized recombination and light in-coupling and EQE of 95% (solid magenta line). The black curve shows the PCE vs. bandgap according to the Shockley-Queisser model.[5] **(c-e)** PCE vs. S and majority-carrier band offsets at both interfaces for the standard cell and the optimized cells with undoped and doped TLs, respectively. Note, a positive offset is defined as energy offset towards midgap, while a negative offset means an energy barrier for electrons and holes.

Based on the bandgap-dependent analysis in **Figure 5**, we propose two different cell architectures employing doped and undoped TLs, which could be realized experimentally to reach the 30% PCE milestone. For the first cell shown in **Figure 6a**, we take advantage of the fact that doping allows increasing the TL thickness without compromising the device performance (**Supplementary Figure S20**, assuming that the parasitic absorption in the TL and the dopant is negligible).[40] This might be an important consideration from the manufacturing perspective as it could help to protect the perovskite from moisture and oxygen ingress. Here, we assumed a bandgap of 3.5 eV for the bottom TL which is comparable to the bandgaps of $TiO_2$ (3.3 eV), $SnO_2$ (3.6 eV) or NiOx.[90,91] Using 100 nm thick TLs, realistic interface recombination velocities of 10 cm/s,[12,22,24] bulk lifetimes of 10 µs (shown in ref.[16]) and doped TLs ($n_{dop}=10^{19}$ $cm^{-3}$),[41] we can simulate a PCE of 30% even for a bandgap of 1.5 eV (which is within reach for FAPI-based perovskites).[32] However, it is clear that realizing a stable doping experimentally without critically enhancing $S$, which may in reality overcompensate the potential benefits (**Figure 5e**) will remain a critical experimental challenge. We note that a thin insulating layer could be implemented to avoid a direct contact between the doped TL and the perovskite potentially lowering the interfacial recombination. The second cell shown in **Figure 6b,** is based on a 1 nm-thick self-assembled monolayer as bottom TL. It was recently shown[92–94] that a self-assembled monolayer (SAM) can replace and even outperform the omnipresent PTAA layer in *p-i-n* type cells with negligible recombination at the bottom interface. Using a SAM at the bottom and a thin TL at the top (20 nm) and the same setting as discussed above for the optimized cell in **Figure 4a**, also allows to reach the 30% milestone; however, we emphasise that this is strictly limited by the thickness of the layer. For example, implementing two 100 nm thick TLs for this particular device reduces the PCE to 27%, which is due to the increased interface loss when lowering the $V_{BI}$ across the perovskite. Nevertheless, we note that the SAM is likely beneficial in terms of parasitic absorption compared to doped TLs and in terms of device stability.[94] The parallel recombination currents in the device for both cells based on the doped TL and the SAM are shown in **Figure 6c**. Finally, by implementing the above quantified ion density ($n_{ion} \sim 6\times10^{16}$ $cm^{-3}$), we observed that the steady-state performance of both proposed high-efficiency cells is hardly affected by the mobile ion density (red dotted lines in **Figure 6a** and **b**). In fact, **Figure 6d** shows that this holds true for ion densities of up to up to $1\times10^{18}$ cm$^{-3}$. This finding can be explained by the increased $V_{BI}$ and reduced $S$ which mitigates the effect of the ion density as previously observed.[68] Therefore, we can conclude, although mobile ions are a key for the perovskite stability they do not represent a significant barrier for perovskite solar cells on their pathway towards the 30% efficiency milestone.

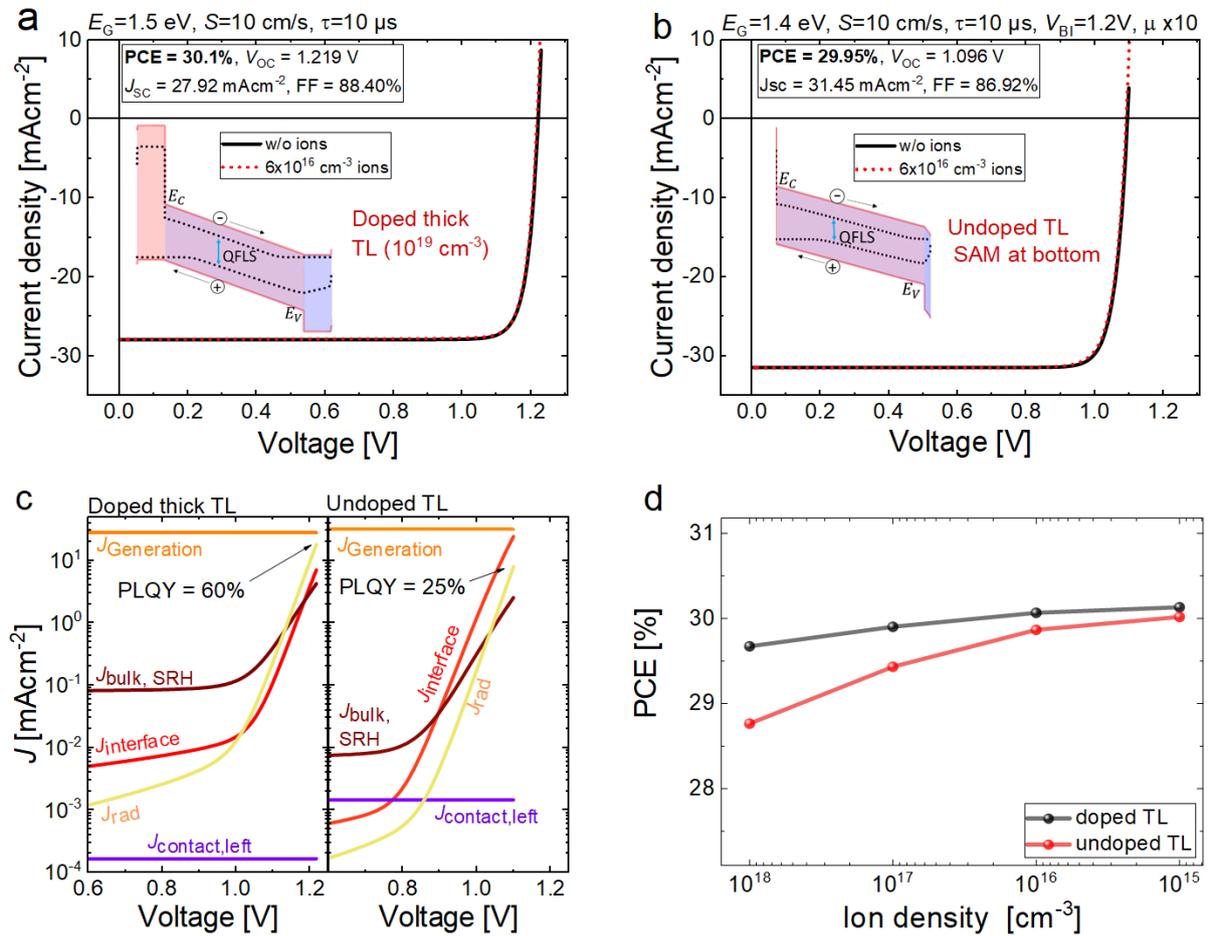

**Figure 6.** (a) Numerically simulated current density vs. voltage (JV) characteristics of p-i-n type cells with comparatively thick, doped TLs ($10^{19}$ cm$^{-3}$) and a bandgap of 1.5 eV, interface recombination velocities (S) of 10 cm/s, a bulk lifetime of 10 μs and a power conversion efficiency (PCE) of ~30%. As shown in this paper, the doped TL maximize the field across the perovskite which enhances the PCE despite significant S values. (b) Numerically simulated JV characteristics of a perovskite cell with a 1 nm-thick self-assembled monolayer at the bottom as an alternative strategy to reach 30% with undoped TLs. Note, that the short-circuit

current and the open-circuit voltage differ between the 2 cells due to their different bandgaps. The red dotted lines in panel (a) and (b) demonstrate that adding the measured ion density of ∼6x10$^{16}$ cm$^{-3}$ while keeping the other simulation parameters identical has an almost negligible effect on the steady-state performance in these 2 high efficiency systems. Panel **(c)** shows the parallel recombination currents as a function of applied voltage in both cells w/o ions. The graph also demonstrates a photoluminescence quantum yield (PLQY) of ∼60% for the cell with doped TLs and ∼25% for the SAM-based cell under open-circuit conditions. This difference is because the built-in potential-drop across the 20 nm thick top TL renders the cell with undoped TLs more vulnerable to interfacial recombination. Therefore, when aiming to avoid doping and keeping the S at a realistic value (S > 1 cm/s), such high PCEs can only be reached in case of ultrathin TLs such as SAMs. **(d)** The steady-state efficiency of the doped-TL cells versus ion density reveals a negligible efficiency loss for ion densities up to 1×10$^{18}$ cm$^{-3}$.

## 3. Conclusion

In summary, in this work we explored possible optimization strategies to advance the PCEs of perovskite solar cells beyond those of monocrystalline silicon (>26%) and GaAs cells (>29%). To this end we established a standard simulation to describe our reference (standard) *p-i-n* type cells with an efficiency of close to 20% using SCAPS and IonMonger to incorporate mobile ions. We then checked a wide range of simulation parameters and discussed the importance of $\mu, n_{\text{ion}}, n_{\text{dop}}, V_{\text{BI}}, E_{\text{maj}}, E_{\text{gap}}, S, \tau_{\text{bulk}}$ as key parameters in defining the cells' performance. In combination with transient capacitance measurements and charge extraction, we demonstrated that the ion density is rather significant (≥6x10$^{16}$ cm$^{-3}$) which significantly redistributes the internal field. Starting from the standard cells, we showed that (10 fold) optimization of the mobilities of the TL and the perovskite layer will result in comparatively marginal PCE improvements (∼1% absolute). Instead, doping the transport layers relaxes the need of electrodes with large work function offsets and enables drastic FF and $V_{\text{OC}}$ gains. We attributed this result to an effective increase in the $V_{\text{BI}}$ across the perovskite layer which drives minority carriers away from the interfaces. We further discussed the need of a high built-in field across all stack layers including the absorber layer, at least at its surfaces. Also, energy level matching between the perovskite and the TL remains a necessity to avoid PCE losses irrespective of the TL doping. We then varied the perovskite bandgap for cells with supressed defect recombination in the bulk and/or the interface. In combination with doped TLs, we identified that recombination parameters that have already been demonstrated in recent literature ($\tau_{bulk}$ = 10 µs and $S$ = 10 cm/s) would allow PCEs of up to 31% for a perovskite with a bandgap of 1.4 eV, if an average above-gap EQE of 95% can be realized. Based on these findings we numerically demonstrated the feasibility of 30% efficient perovskite cells with comparatively thick (≈ 100 nm), doped TLs (~$10^{19}$ cm$^{-3}$) by implementing a perovskite layer with realistic interface recombination velocities ($S$ = 10 cm/s) and a bandgap of 1.5 eV. Alternatively, cells with ultrathin TLs (e.g. by implementing self-assembled monolayers at the bottom contact) may also allow to reach the 30% milestone, if the $V_{BI}$ across the perovskite layer can be maximized to supress the critical interface recombination. However, although doped TLs allow to relax several strict requirements to realize such near-ideal solar cells; e.g. by allowing to use thicker TLs and by increasing the tolerance against interfacial defects; SAMs are beneficial in terms of optical losses and device stability, which is a key for future developments. Finally, we also demonstrated that the rather high ion density does not represent a major roadblock to unlock efficiencies at the radiative limits and therefore, we expect that perovskite solar cells will become the most efficient single-junction solar technology in the near future.

**Supporting Information**

Supporting Information is available from the Wiley Online Library or from the author.

**Acknowledgements**

This work was funded by the Deutsche Forschungsgemeinschaft (DFG, German Research Foundation) - project numbers 423749265 and 424216076 as well as by HyPerCells (a joint graduate school of the Potsdam University and the HZB). The work of M.H.F. and B.E. is part of the Dutch Research

Council (NWO) and was performed at the research institute AMOLF. The work by V.M.L.C. is supported by a grant from STW/NWO (VIDI 13476). This is a publication by the FOM Focus Group \Next Generation Organic Photovoltaics", participating in the Dutch Institute for Fundamental Energy Research (DIFFER). TK acknowledge the Helmholtz Association for funding via the PEROSEED project. We also thank Nicola Courtier for her kind support with IonMonger.

Received: ((will be filled in by the editorial staff))
Revised: ((will be filled in by the editorial staff))
Published online: ((will be filled in by the editorial staff))

Jonas Diekmann, Pietro Caprioglio, Moritz H. Futscher, Vincent M. Le Corre, Sebastian Reichert, Frank Jaiser, Malavika Arvind, Lorena Perdigón Toro, Emilio Gutierrez-Partida, Francisco Peña-Camargo, Carsten Deibel, Bruno Ehrler, Thomas Unold, Thomas Kirchartz, Dieter Neher, Martin Stolterfoht[*]**Pathways towards 30% efficient single-junction perovskite solar cells and the role of mobile ions**

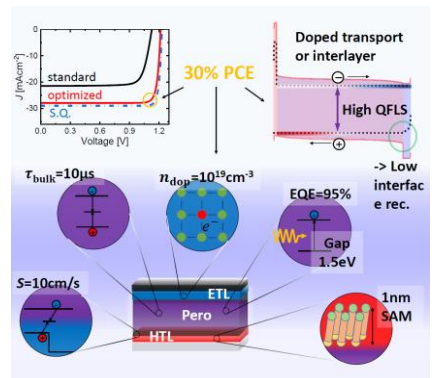

Here, we establish drift-diffusion simulation parameters to describe efficient (20%) *p-i-n* type perovskite solar cells. Using this parameter set, we explore effective strategies to improve the performance further. We find that the key to reach the 30% efficiency milestone is maximizing the built-in voltage across the perovskite layer by implementing doped- or ultrathin transport layers like self-assembled monolayers.

**Pathways towards 30% efficient single-junction perovskite solar cells and the role of mobile ions**

*Jonas Diekmann, Pietro Caprioglio, Moritz H. Futscher, Vincent M. Le Corre, Sebastian Reichert, Frank Jaiser, Malavika Arvind, Lorena Perdigón Toro, Emilio Gutierrez-Partida, Francisco Peña-Camargo, Carsten Deibel, Bruno Ehrler, Thomas Unold, Thomas Kirchartz, Dieter Neher, Martin Stolterfoht[*]*

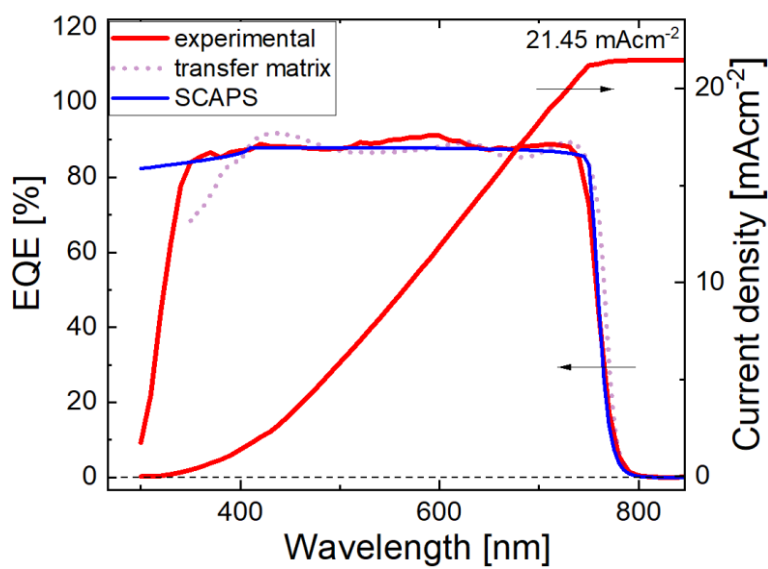

**Supplementary Figure S1.** Experimental External Quantum Efficiency of the standard cells and numerically simulated EQEs using SCAPS and on optimal transfer matrix simulations providing an integrated $J_{SC}$ of roughly 21.5 mAcm$^{-2}$.

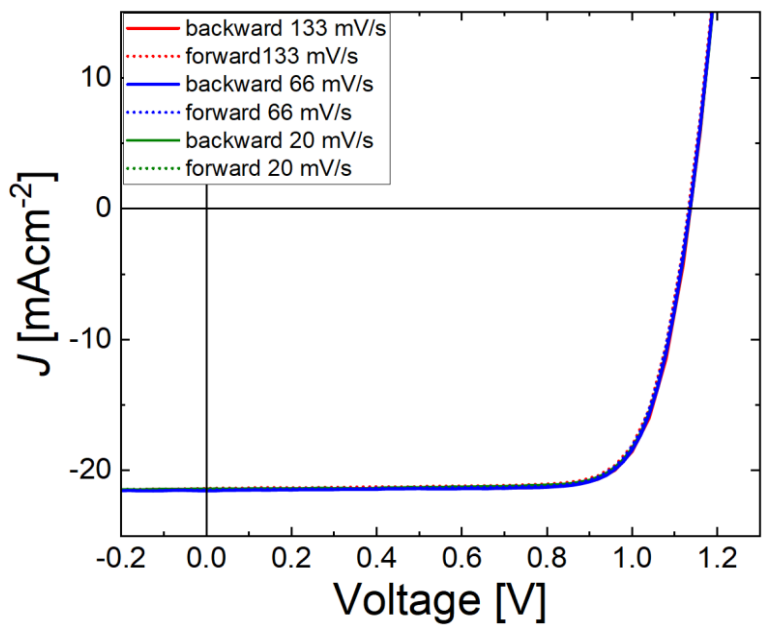

**Supplementary Figure S2.** JV-scans of the standard cell measured in forward and reverse direction at different scan rates demonstrating a negligible hysteresis within the studied time-range accessible in our standard JV measurement setup using a Keithley 2400.

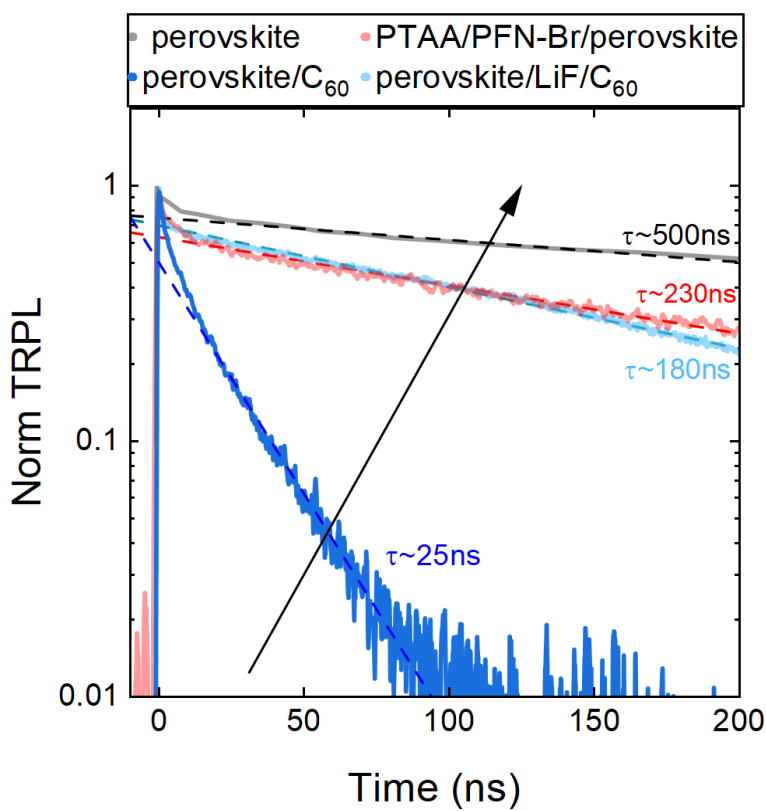

**Supplementary Figure S3.** Transient PL measurements on partial triple cation cell stacks with and without transport layers (PTAA/PFN-Br and (LiF/)$C_{60}$). Using the expression shown in ref.[1] $\left(\frac{1}{\tau_{eff}} = \frac{1}{\tau_{bulk}} + \left(\frac{4d^2}{\pi^2 D} + \frac{d}{S}\right)^{-1}\right)$, where $\tau_{eff}$ and $\tau_{bulk}$ is the effective and bulk lifetime respectively, d the film thickness (400 nm), S the interface recombination velocity, $D = \mu V_{th}$ is the diffusion coefficient, which is the product of mobility and thermal voltage (25.86 meV at 300 K), we can quantify the interface recombination velocity $S = \frac{dD\pi^2(\tau_{bulk}-\tau)}{(D\pi^2\tau_{bulk}\tau_{eff} - 4d^2(\tau_{bulk}-\tau))}$ to be approximately 2000 cm/s at the perovskite/$C_{60}$ interface, and 100 cm/s at the PTAA/PFN-Br/perovskite interface. We note, the improvement in lifetime upon addition of LiF at the n-interface is also shown, which we have used to model the cell shown in **Figure 1e**.

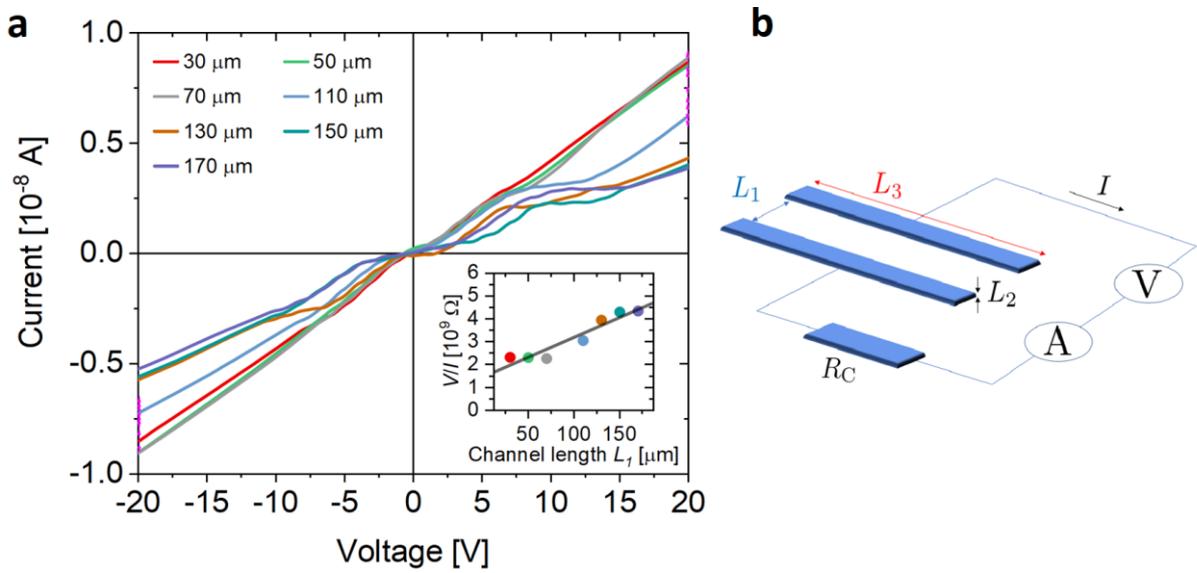

**Supplementary Figure S4.** (**a**) Current density versus voltage characteristics measured on a transistor-like structure shown in (**b**) to measure the perovskite conductivity. The electrode width ($L_3$) was 1 cm, the perovskite thickness ($L_2$) was 500 nm, $R_c$ refers to the contact resistance. The channel length $L_1$ is variable. The conductivity of $\sigma \sim 1 \cdot 10^{-7}$ S/cm is obtained from a fit of the slope of resistance versus channel length using the expression $\frac{V}{I} = \frac{1}{\sigma}\frac{L_1}{L_2 L_3} + R_c$. Considering that the conductivity is $\sigma = en\mu$ and using a mobility of 1 cm$^2$/Vs yields an equilibrium carrier density of $6 \times 10^{11}$ cm$^{-3}$. We note any equilibrium carrier density below $1 \times 10^{14}$ cm$^{-3}$ has no effect on the simulation results.

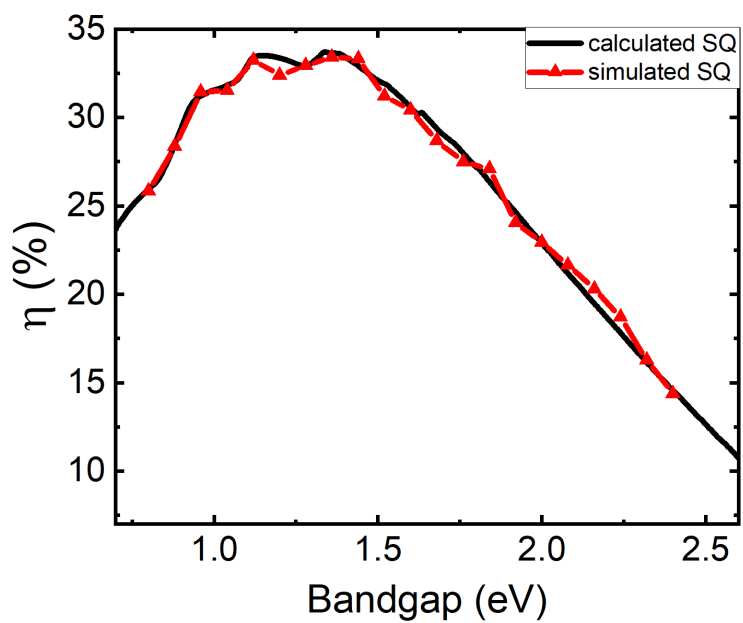

**Supplementary Figure S5.** A comparison between the analytic Shockley-Queisser model (assuming a perfect rear reflector)[2] and numerical simulations assuming only radiate recombination losses using a bandgap-dependent $k_2$ as described below in **Supplementary Note 1**.

**Supplementary Table S1.** SCAPS simulation parameters for the standard cells.

| Parameter | Symbol | Value | Unit | Comment |
|---|---|---|---|---|
| Majority carrier band offset between perovskite and $C_{60}$ | $\Delta E_{maj,c}$ | 0 | eV | Due to QFLS=$V_{oc}$ match in these devices, ref.[3] |
| Majority carrier band offset between perovskite and PTAA | $\Delta E_{maj,v}$ | 0 | eV | Due to QFLS=$V_{oc}$ match in these devices and UPS data, ref.[3] |
| Lifetime in perovskite | $\tau_{pero}$ | 500 | ns | Measured with TRPL, Figure S3 |
| Trap density perovskite | $N_{t,pero}$ | $2\cdot 10^{15}$ | $cm^{-3}$ | From the measurement of the lifetime, see caption* |
| Lifetime in PTAA | $\tau_p$ | 1 | ns | Sub-ns exciton lifetime in organic conjugated polymer, ref.[4] |
| Lifetime in $C_{60}$ | $\tau_n$ | 1 | ns | Sub-ns exciton lifetime in organic conjugated polymer, ref.[5] |
| Ionized acceptors in PTAA | $N^-_{A,p}$ | 0 | $cm^{-3}$ | No doping used in fabrication |
| Ionized donors in $C_{60}$ | $N^+_{D,n}$ | 0 | $cm^{-3}$ | No doping used in fabrication |
| Intrinsic carrier density in perovskite | $N_i$ | $1\times 10^{12}$ | $cm^{-3}$ | Effectively undoped (irrelevant <$1\times 10^{14}$ $cm^{-3}$) as evidenced by the low conductivity, Figure S4 |
| Minority carrier recombination velocity from perovskite to PTAA | $S_{min,n}$ | 100 | cm/s | Measured with TRPL, Figure S3 |
| Trap density at $p$-interface | $N_{t,p-if}$ | $1\cdot 10^{11}$ | $cm^{-2}$ | From the measurement of interface rec. velocity, see caption** |

| Parameter | Symbol | Value | Unit | Source |
|---|---|---|---|---|
| Minority carrier recombination velocity from perovskite to C$_{60}$ | $S_{\text{min},p}$ | 2000 | cm/s | Measured with TRPL, Figure S3 |
| Trap density at *n*-interface | $N_{\text{t},n-\text{if}}$ | 2·10$^{12}$ | cm$^{-2}$ | From the measurement of interface rec. velocity, see caption** |
| Majority carrier recombination velocity from perovskite to PTAA | $S_{\text{maj},p}$ | 1x10$^7$ | cm/s | Assumption |
| Majority carrier recombination velocity from perovskite to C$_{60}$ | $S_{\text{maj},n}$ | 1x10$^7$ | cm/s | Assumption |
| Majority and Minority carrier velocity at front metal contact (ITO) | $S_{\text{met}}$ | 1x10$^7$ | cm/s | Typical value for metal, ref.[6] |
| Majority and Minority carrier velocity at back metal contact (Cu) | $S_{\text{met}}$ | 1x10$^7$ | cm/s | Typical value for metal, ref.[6] |
| Thickness of PTAA | $d_{\text{PTAA}}$ | 10 | nm | Measured with profilometer |
| Thickness of perovskite | $d_{\text{pero}}$ | 400 | nm | Measured with profilometer |
| Thickness of C$_{60}$ | $d_{\text{C60}}$ | 30 | nm | Measured with profilometer |
| Offset between metal and PTAA | $\Delta E_{\text{F,metal}-}$ | 0.2 | eV | Assumption |
| Offset between metal and C$_{60}$ | $\Delta E_{\text{F,metal}-}$ | 0.2 | eV | Assumption |
| Device built-in voltage | $V_{\text{BI}}$ | 1.2 | V | Fit to experimental data, see caption*** |
| Bandgap PTAA | $E_{\text{G,PTAA}}$ | 3.0 | eV | Measured by UV-Vis |
| Electron affinity PTAA | $E_{\text{A,PTAA}}$ | 2.5 | eV | Consider aligned PTAA HOMO with perovskite valence band |
| Bandgap perovskite | $E_{\text{G,pero}}$ | 1.63 | eV | Measured by UV-Vis |
| Electron affinity perovskite | $E_{\text{A,pero}}$ | 3.9 | eV | Using photon yield spectroscopy measurements, ref.[3] |
| Electron affinity C$_{60}$ | $E_{\text{A,C60}}$ | 3.9 | eV | Consider aligned C$_{60}$ LUMO with perovskite conduction band |
| Bandgap C$_{60}$ | $E_{\text{G,C60}}$ | 2.0 | eV | Measured by UV-Vis |
| Electron mobility in C$_{60}$ | $\mu_{n,\text{C60}}$ | 1x10$^{-2}$ | cm$^2$/Vs | 1.5·10$^{-2}$ cm$^2$/Vs measured with SCLC in ref.[7] Ref.[8] showed that $\mu_{n,\text{C60}}$ and $\mu_{n,\text{pero}}$ are much larger than $\mu_{p,\text{PTAA}}$ in these devices. |
| Hole mobility in PTAA | $\mu_{p,\text{PTAA}}$ | 1.5x10$^{-4}$ | cm$^2$/Vs | Fit to thickness dependent PTAA *JV*-curves (Figure 1g). A mobility of 7.5·10$^{-5}$ cm$^2$/Vs measured on undoped PTAA with SCLC in ref.[9] |
| Electron mobility in perovskite | $\mu_{n,\text{pero}}$ | 1 | cm$^2$/Vs | Intra-grain Thz perovskite mobilities are around 35 cm$^2$/Vs for triple cation perovskites.[10,11] However, a better fit to the *JV*-curves shown in Figure 1 was obtained with a mobility of 1 |
| Hole mobility in perovskite | $\mu_{p,\text{pero}}$ | 1 | cm$^2$/Vs | |
| | | | | cm$^2$/Vs, which might be related to grain boundaries. |
| Relative dielectric constant PTAA | $\epsilon_{PTAA}$ | 3.5 | | Typical value for organic conjugated polymers.[12] |
| Relative dielectric constant perovskite | $\epsilon_{pero}$ | 22 | | Typical value, e.g. $\epsilon_{pero}$ = 25 in ref.[13] |
| relative dielectric constant C$_{60}$ | $\epsilon_{C60}$ | 5.0 | | a relative permittivity of ~5 was measured at lowest frequencies (1kHz) at 300 K, ref.[14] |
| Effective electron density of states in HTL | $N_{C/V,PTAA}$ | 1x10$^{20}$ | cm$^{-3}$ | Ref.[15] |
| Effective electron density of states in C$_{60}$ | $N_{C/V,C60}$ | 1x10$^{20}$ | cm$^{-3}$ | Ref.[15] |
| Effective electron density of states in perovskite | $N_{C/V,pero}$ | 2.2x10$^{18}$ | cm$^{-3}$ | Ref.[16] |

Remarks: We note that we do not claim that every individual parameter is perfectly consistent with existing literature, however, we consider them as plausible. Many parameters were obtained from our own optical and electro-optical transient measurements, such as the recombination rate constants and lifetimes in the perovskite layer, interfacial recombination velocities, energy levels and density of states in the perovskite. The absorption coefficient (α [1/cm]) of the perovskite layer in the simulations was matched to the experimentally measured absorption coefficient, in particular at the band edge which has a significant impact on the EQE and short-circuit current. We further implemented an Urbach tail of 15 meV consistent with experimental data. *From the measured lifetime (**Supplementary Figure S3**) in the perovskite ($\tau_{pero} = 1/(N_{t,pero}v_{th}\sigma)$), a trap state density of $2·10^{15}$ cm$^{-3}$ can be estimated by assuming a capture cross section of $10^{-16}$ cm$^{-2}$ and a thermal velocity of 1x10$^7$ cm/s. **From the obtained interface recombination velocities (**Supplementary Figure S3**), S= ($N_{t,if}v_{th}\sigma$), a trap state density of $1·10^{11}$ cm$^{-2}$ (S=100 cm/s) and $2·10^{12}$ cm$^{-2}$ (S=2000 cm/s) can be estimated at the p-interface and at the n-interface, respectively, by assuming a capture cross section of $10^{-16}$ cm$^{-2}$ and a thermal velocity of 1x10$^7$ cm/s. ***Figure 4a** shows that a high V$_{BI}$ of ~1.2 V is necessary in case of undoped transport layers and the given interface recombination velocities to achieve the measured V$_{OC}$. This is consistent with a large surface photovoltage of approximate 1V obtained on ITO/PTAA/PFN-Br/triple cation perovskite stacks with UPS.[17]

**Supplementary Note S1.**

For simulating the cells without ions, we used the well-established software SCAPS.[18,19] SCAPS numerically solves a system of three coupled equations, namely the Poisson equation, the continuity equation and the drift-diffusion equation. The Poisson equation is given by

$$\frac{\partial^2 \varphi(x)}{\partial x^2} = -\frac{q}{\epsilon}(p(x) - n(x) + N_D^+ - N_A^-) \qquad \text{(S1)}$$

where φ is the electrical potential, $q$ is the charge of an electron, $\epsilon$ is the absolute permittivity of the material, $p(x)$ and $n(x)$ are the the hole and electron density respectively and $N_D^+$ and $N_A^-$ are the Donor and the Acceptor densities. The continuity equations for the electrons and holes are given by

$$\frac{\partial n}{\partial t} = -\frac{1}{q}\frac{\partial J_n}{\partial x} - R(x) + G(x) \qquad \text{(S2a)}$$

$$\frac{\partial p}{\partial t} = -\frac{1}{q}\frac{\partial J_p}{\partial x} - R(x) + G(x) \qquad \text{(S2b)}$$

where $J_n$ and $J_p$ are the electron and hole current densities respectively, $R(x)$ is the recombination rate and $G(x)$ is the generation rate. The drift-diffusion equations are given by

$$J_n = qn(x)\mu_n \frac{\partial \varphi}{\partial x} + qD_n \frac{\partial n(x)}{\partial x} \qquad \text{(S3a)}$$

$$J_p = qp(x)\mu_p \frac{\partial \varphi}{\partial x} + qD_p \frac{\partial p(x)}{\partial x} \qquad \text{(S3b)}$$

where $\mu_n$ and $\mu_p$ are the mobilities of electrons and holes, $D_n$ and $D_p$ are the diffusion constants of electrons and holes. The mobility and the diffusion constants are related via the Einstein relation

$$D = \mu k_\text{B} T/q \qquad \text{(S4)}$$

The surface recombination at the metal electrodes dictates the boundary conditions of the carrier current densities for the cell as a whole

$$J_\text{n}(-d_\text{H}) = qS_\text{n,fc}\bigl(n(-d_\text{H}) - n_\text{o}(-d_\text{H})\bigr) \qquad \text{(S5a)}$$

$$J_\text{p}(-d_\text{H}) = qS_\text{p,fc}\bigl(p(-d_\text{H}) - p_\text{o}(-d_\text{H})\bigr) \qquad \text{(S5b)}$$

$$J_\text{n}(d + d_\text{E}) = qS_\text{n,bc}\bigl(n(d + d_\text{E}) - n_\text{o}(d + d_\text{E})\bigr) \qquad \text{(S5c)}$$

$$J_\text{p}(d + d_\text{E}) = qS_\text{p,bc}\bigl(p(d + d_\text{E}) - n_\text{o}(d + d_\text{E})\bigr) \qquad \text{(S5d)}$$

Here, $S$ is the recombination velocity which is given by $S = N_\text{t,if} v_\text{th} \sigma$, where $N_\text{t,if}$ is the trap density at the interface, $v_\text{th}$ the thermal velocity and $\sigma$ the capture cross section. Fc, bc denote the front contact and back contact. $d_E$, $d_H$ and $d$ are the thicknesses of the of the electron-transport layer, the hole transport layer and the active layer respectively. $n_0$ and $p_0$ are the equilibrium electron and hole carrier densities. The equilibrium majority carrier concentration at a metal-semiconductor interface is related to the difference of the Fermi-levels, also known as the Schottky barrier height $\varphi_b$ as follows

$$p_0(x = -d_\text{H}) = N_\text{V}\, exp\left(\frac{-\varphi_\text{b}(-d_\text{H})}{k_\text{B}T}\right) \qquad \text{(S6a)}$$

$$n_0(x = d + d_E) = N_C \, exp\left(\frac{-\varphi_b(d + d_E)}{k_B T}\right) \quad \text{(S6b)}$$

where $N_C$ and $N_V$ denote the effective density of states in the conduction and valence band respectively. The minority carrier concentration at equilibrium can then be calculated from these results using the following equation

$$n_0(x)p_0(x) = n_i^2 = N_C N_V \, exp\left(\frac{-E_g}{k_B T}\right) \quad \text{(S7)}$$

where $n_i$ is the intrinsic carrier concentration and $E_g$ the bandgap.

Similarly, the interface recombination dictates the boundary conditions at the perovskite-ETL (back interface, bi) and HTL-perovskite (front interface, fi) interfaces

$$J_n(0) = qS_{n,fi}(n(0) - n_o(0)) \quad \text{(S8a)}$$

$$J_p(0) = qS_{p,fi}(p(0) - p_o(0)) \quad \text{(S8b)}$$

$$J_n(d) = qS_{n,bi}(n(d) - n_o(d)) \quad \text{(S8c)}$$

$$J_p(d) = qS_{p,bi}(p(d) - n_o(d)) \quad \text{(S8d)}$$

The Recombination in the bulk material is given by the following equations

$$R(x) = R_{\text{SRH}} + R_{\text{radiative}} + R_{\text{Auger}} \tag{S9a}$$

$$R_{\text{radiative}}(x) = k_2(n(x)p(x) - n_i^2) \tag{S9b}$$

$$R_{\text{Auger}}(x) = (c_n^A n + c_p^A p)(n(x)p(x) - n_i^2) \tag{S9c}$$

$$R_{\text{SRH}} = \frac{n(x)p(x) - n_i^2}{(n(x) + n_1)\tau_p + (p(x) + p_1)\tau_n} \tag{S9d}$$

where $k_2$ is the radiative radiation coefficient, $c_n^A$ and $c_p^A$ are the Auger coeffients for electrons and holes respectively, $\tau_p$ and $\tau_n$ are the SRH-lifetimes for holes and electrons and $n_1 = N_C \exp(-(E_C - E_T)/k_B T)$ and $p_1 = N_V \exp((E_V - E_T)/k_B T)$ is the generation of electrons in the conduction band and holes in the valence band from the trap level, respectively. The minimum SRH lifetime in the bulk is given by

$$\tau_{p/n} = 1/(N_t v_{\text{th}} \sigma) \tag{S10}$$

where $N_t$ is the trap density in the bulk. The Generation is given by

$$G(x) = \int_{\lambda_{\min}}^{\lambda_{\max}} \alpha(\lambda) N_{\text{phot}}(\lambda, x) d\lambda \tag{S11}$$

where α is the wavelength dependent absorptivity, which we determined experimentally, $N_{\text{phot}}$ is the photon flux at each position, which is given by

$$N_{\text{phot}}(\lambda, x) \tag{S12}$$
$$= N_{\text{phot},0}(\lambda) T_{\text{front}} \exp(-x\alpha(\lambda)) \frac{1 + R_{\text{back}} \exp(-2(d-x)\alpha(\lambda))}{1 - R_{\text{back}} R_{\text{int}} \exp(-2d\alpha(\lambda))}$$

where $N_{phot,0}(\lambda)$ is the incident photon flux, in this case the AM1.5G spectrum, $T_{front}$ is the transmission of the front contact, $R_{int}$ is the internal reflection at the front contact and $R_{back}$ is the reflection of the back contact.

**Supplementary Note S2.**

We obtain the external bimolecular recombination rate constant $k_2$ from the radiative recombination current density in the dark ($J_{0,rad}$) and knowledge of the effective density of states $N_C = N_V = 2.2 \times 10^{-18}$ cm$^{-3}$ (ref.[20]) which corresponds to an effective electron mass $m_e^*$ of 0.2.

$$J_{0,rad} = edk_2 n_i^2 = edk_2 N_C N_V exp(-E_G/k_B T), \qquad (S13)$$

where $e$ is the elementary charge, $d$ is the film thickness, $E_G$ the bandgap and $k_B T$ the thermal energy. Note, $J_{0,rad}$ is obtained from integrating the EQE and black body spectrum ($\phi_{BB}$)

$$J_{0,rad} = \int_0^\infty EQE \cdot \phi_{BB}\, dE \qquad (S14)$$

Considering that $\phi_{BB} \propto E_G^2 exp(-E_G/k_B T)$ and $J_{0,rad} \propto exp(-E_G/k_B T)$ shows that $k_2$ must depend on the bandgap itself. For a bandgap of 1.63 eV and $d$ = 500 nm, we obtain an external $k_2$ of $3 \times 10^{-11}$ cm$^3$/s which is similar to previously specified values ref.[20–22]

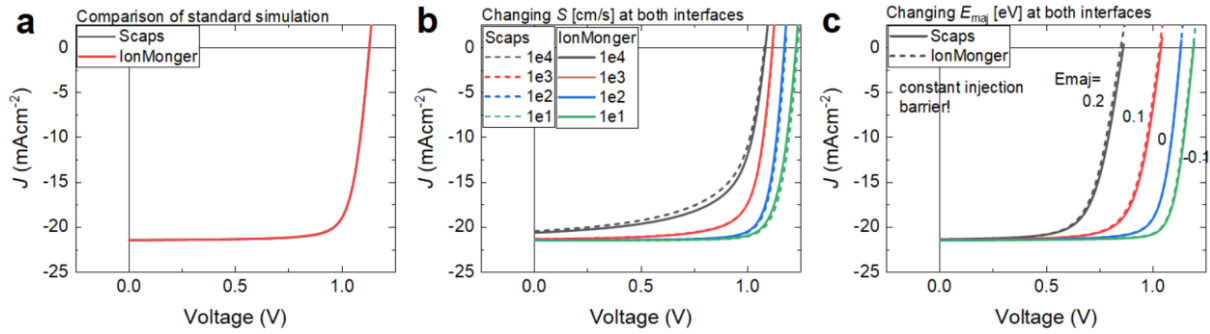

**Supplementary Figure S6**. **a** A direct comparison between IonMonger and SCAPS using the standard simulations settings in **Supplementary Table S1** without mobile ions. These settings could well reproduce the standard simulation setting of SCAPS. In order to check whether both programs lead to identical results with respect to key recombination parameters, such as the interface recombination velocity (S) and the majority carrier offset ($E_{maj}$), we varied these 2 parameters as shown in panel **b** and **c**, which resulted in very similar results.

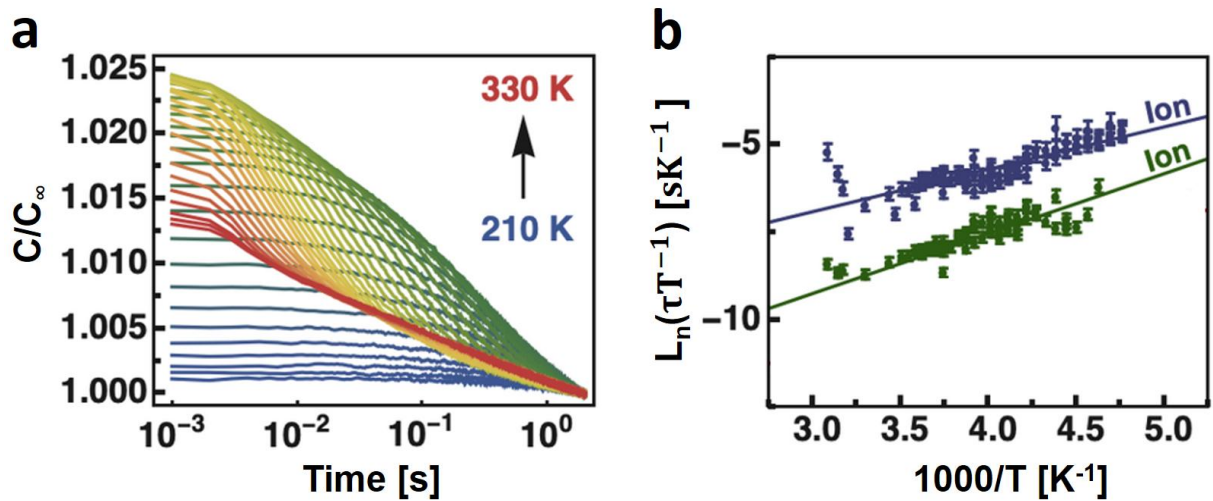

**Supplementary Figure S7**. Transient capacitance measurements. **a** Change of capacitance as a function of time after switching from 1.25 V (applied for 2 seconds) to 0 V measured between 210 and 330 K in steps of 3 K. The capacitance is measured at 0 V by applying a sinusoidal voltage with an AC amplitude of 20 mV to the device at high frequencies ($10^4$ Hz) where the measured capacitance corresponds to the geometric capacitance of the device. The change in capacitance directly shows the redistribution of mobile ions in the internal field under short-circuit conditions. The measured capacitance is divided by the steady-state capacitance, which is 56 nF/cm$^2$ at 300 K. **b** Arrhenius plot of the obtained capacitance decay times for the filling voltages shown in b. Fits with a confidence level below p = 0.2 are rejected. The linear fit reveals the characteristics for two mobile ion species (Ion 1: Activation energy = 121 ± 1 meV, diffusion coefficient = (1.1 ± 0.1) x $10^{-10}$ cm$^2$/s and Ion 2: Activation energy = 118 ± 1 meV, diffusion coefficient = (5.8 ± 3.1) x $10^{-10}$ cm$^2$/s). Diffusion coefficients are calculated at 300 K. The values were obtained assuming a perovskite thickness of 400 nm, a fully depleted perovskite layer with a built-in voltage of 1.0 V, that diffusion of mobile ions is negligible against drift, and that the electric field varies linearly within the perovskite layer. The obtained

time constants can then be approximated as $\tau = \frac{d_{pero}^2 k_B T}{D q V_{BI}}$. The diffusion coefficient is given by $D = D_0 \exp(-E_A/k_B T)$, where $D_0$ is a pre-factor and $E_A$ the activation energy.

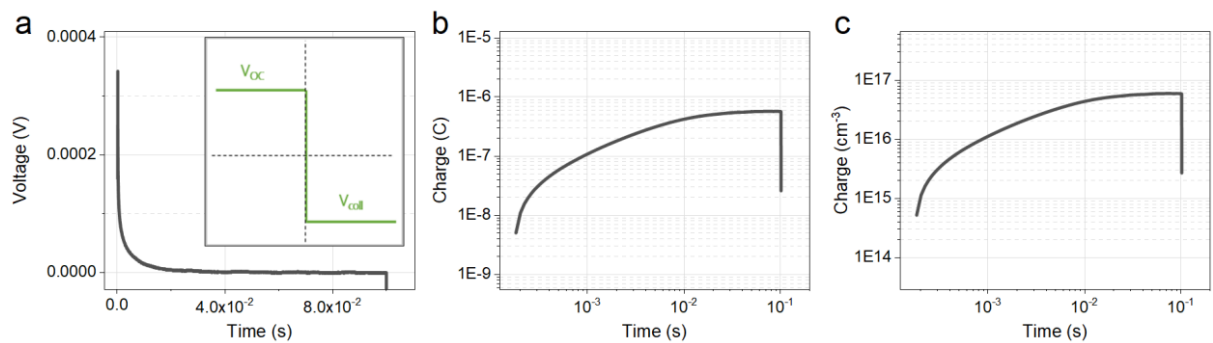

**Supplementary Figure S8. a** Bias assisted transient charge extraction measurements are performed by switching the voltage from open-circuit conditions to a reverse bias of -1.6 V to collect the charges (collection bias). It is important to note that the extraction of injected charges (during the application of the $V_{OC}$ and the RC-charging of the electrodes happens on much faster times scales, i.e. <5 µs in the studied device), allowing to rule out the contribution of these 2 processes on the extracted charge on these long times scales presented here. The extracted charge in panel **b** shows that the ions are transported through the active layer on timescales of around tens of ms which roughly matches the diffusion constants calculated from transient capacitance measurements. **c** By assuming a homogenous distribution of ions in the active layer under open-circuit conditions, we can estimate the ionic charge to be at least $6 \times 10^{16}$ cm$^{-3}$. However, as larger reverse biases consistently break our devices, larger ion densities cannot be excluded at this point.

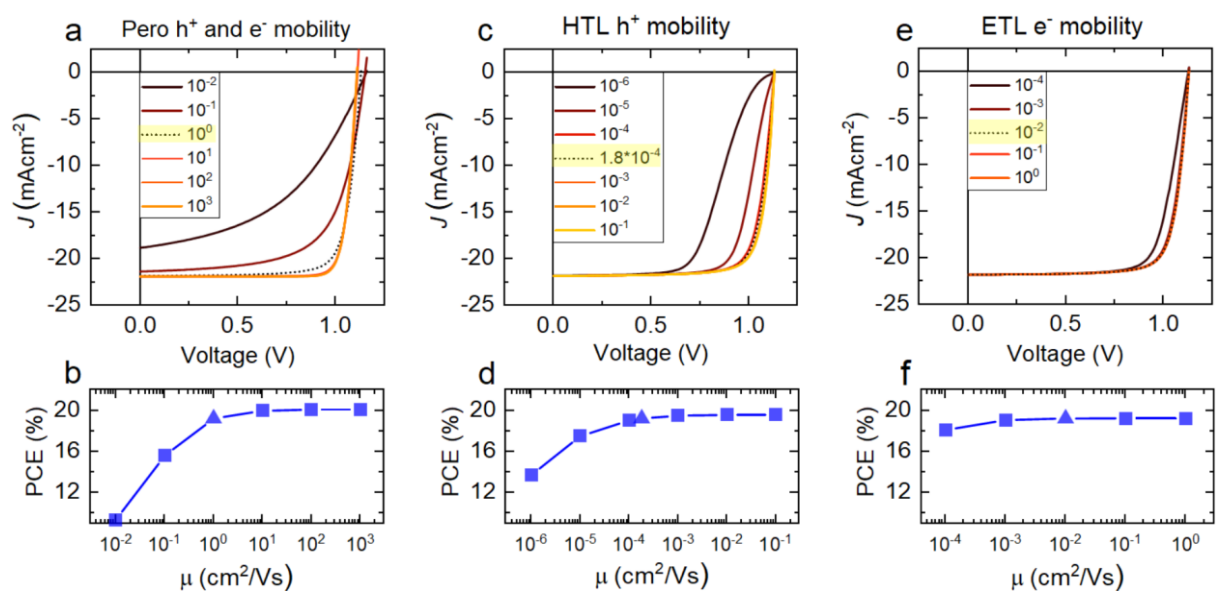

**Supplementary Figure S9.** Numerically simulated current density vs. voltage characteristics of p-i-n type perovskite solar cells by varying the e$^-$ and h$^+$ mobility ($\mu_{e-}$, $\mu_{h+}$) in the perovskite layer **(a)** and resulting power conversion efficiency (PCE) vs. perovskite mobility **(b)**; $\mu_{h+}$ in the HTL **(c)** and resulting PCE vs. HTL mobility **(d)**, $\mu_{e^-}$ in the ETL **(e)** and resulting PCE vs. ETL mobility **(f)**.

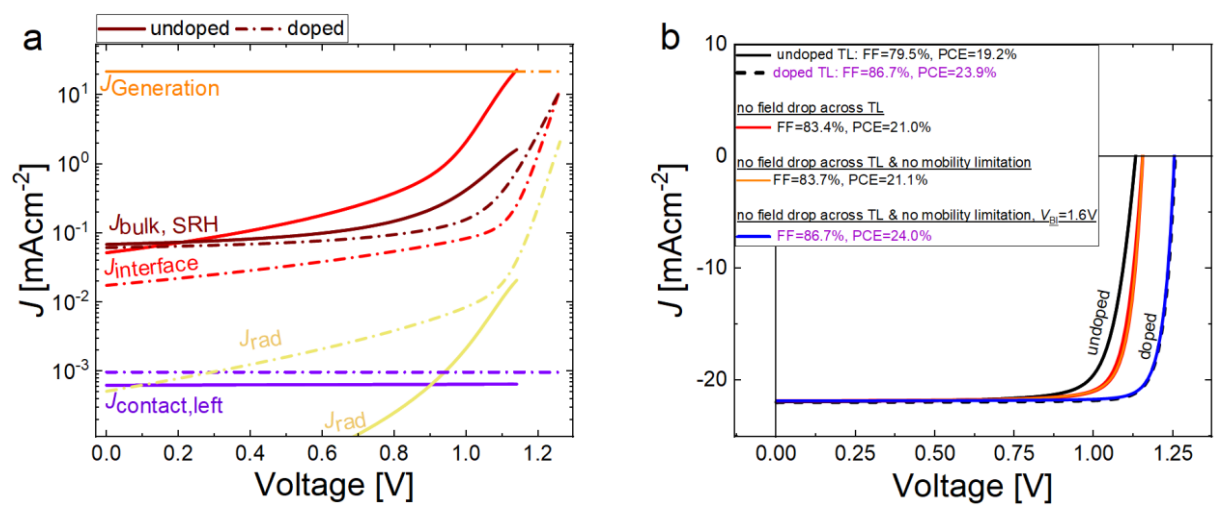

**Supplementary Figure S10.** **(a)** A comparison of the parallel interfacial recombination currents for a simulated standard cell with (dashed) and without (solid) doped TLs. The main effect of the doping is the suppression of the interfacial recombination current (red) which is significantly lower than in the reference cell (e.g. ~25x at an applied voltage of 1.0 V). **(b)** compares the JV-curves of cells with and without doped TLs to a cell with no $V_{BI}$ drop across the TLs (red). This is achieved by including a high relative dielectric constant (e.g. $\epsilon_r$=1000) in the TL. Further optimizing the TL mobilities (100x higher compared to the reference cell) shows that the transport limitations are negligible (orange line). Further maximizing the $V_{BI}$ to 1.6 eV (blue) can closely describe the cell with doped TLs (where the small differences dependent on the exact value of $\epsilon_r$). This comparison shows that the effective increase of the $V_{BI}$ across the perovskite absorber layer due to the doping of the TLs is the primary reason for the performance improvement.

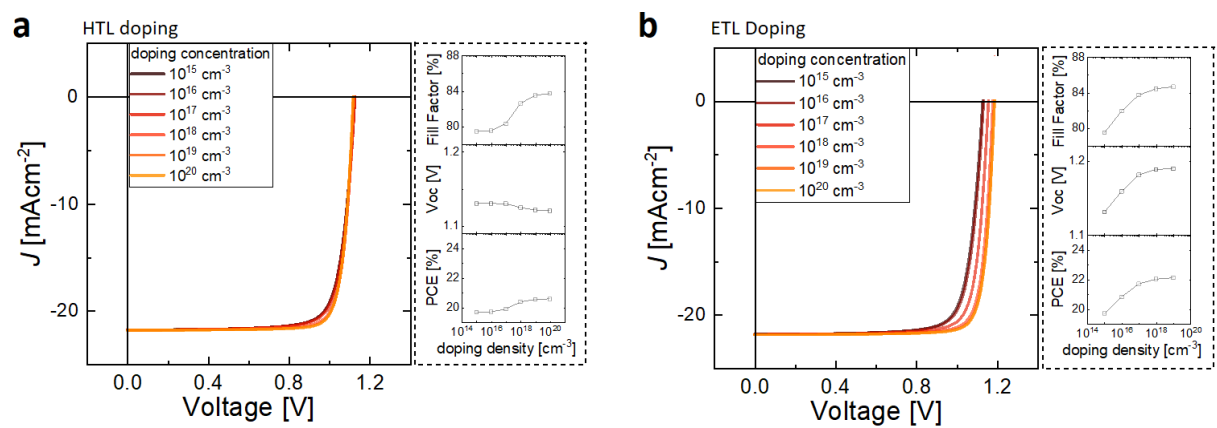

**Supplementary Figure S11.** Simulated JV characteristics of p-i-n type perovskite solar cells by varying the acceptor concentration in the hole transport layer (TL) **(a)** and the donor concentration in the electron TL **(b)**. Considering that perovskite/ETL interface is limiting the performance, doping the HTL does allow performance improvements. In contrast significant

improvements are achieved upon doping of the ETL until the HTL/perovskite interface and or the bulk limit the cells performance.

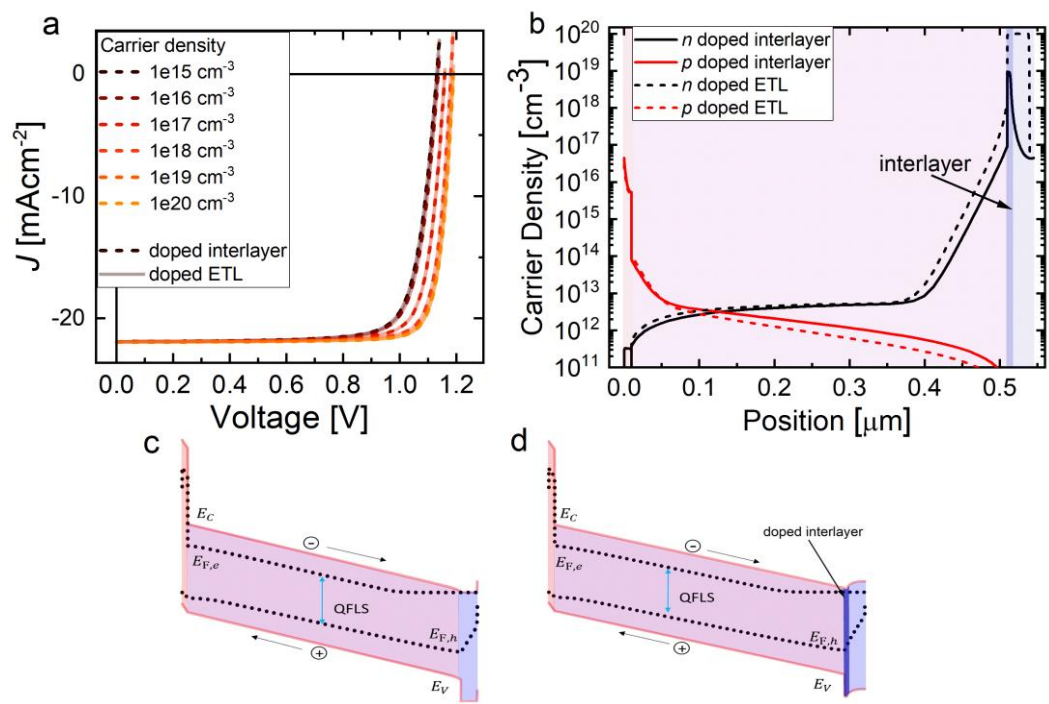

**Supplementary Figure S12.** **(a)** Simulated JV characteristics of p-i-n type perovskite solar cells for different donor concentrations in the electron transport layer (ETL, solid lines) compared to varying the doping in an interlayer between the perovskite and the ETL (dashed lines). Both possibilities yield the same PCE, therefore enhancing the experimental possibilities to reduce the critical interfacial recombination at the top-interface through implementation of a doped layer. **(b)** A comparison of the electron and hole density in the cell in case of the doped ETL and the doped interlayer for a donor concentration of $10^{19}$ cm$^{-3}$. A doped interlayer (here implemented between the perovskite and the ETL) gives the same PCE as a doped ETL. Panel **(c)** and **(d)** demonstrate the simulated band diagram in case of a doped ETL and interlayer, respectively.

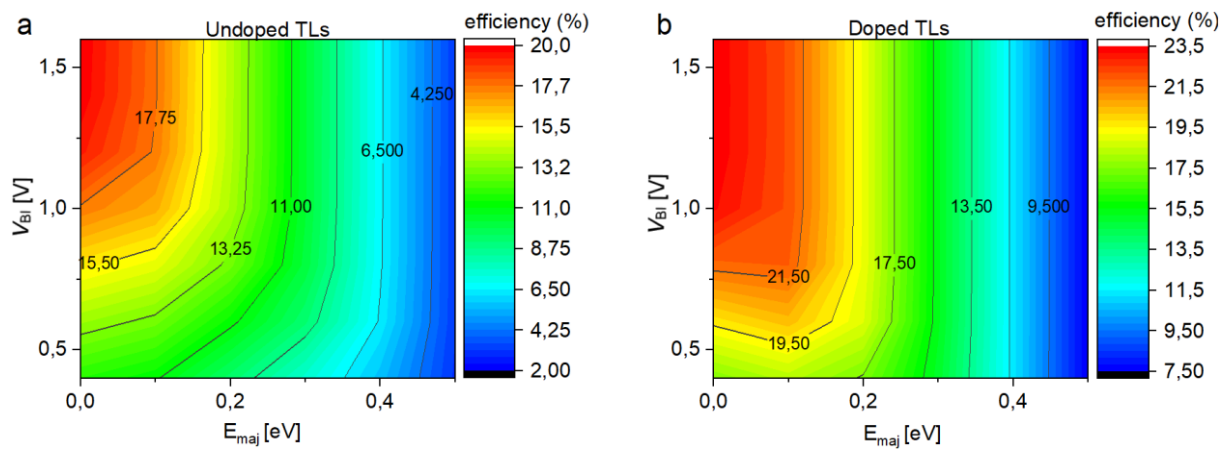

**Supplementary Figure S13.** (a) Heat maps of PCE vs. built-in voltage and majority offset ($E_{maj}$) at both ETL and HTL for the standard cell with undoped transport layers (a) and for the same cell with doped TLs (b). Doped TLs allow more tolerance with respect to the electrode work function difference (defined as the $V_{BI}$ in the work) but not with respect to $E_{maj}$.

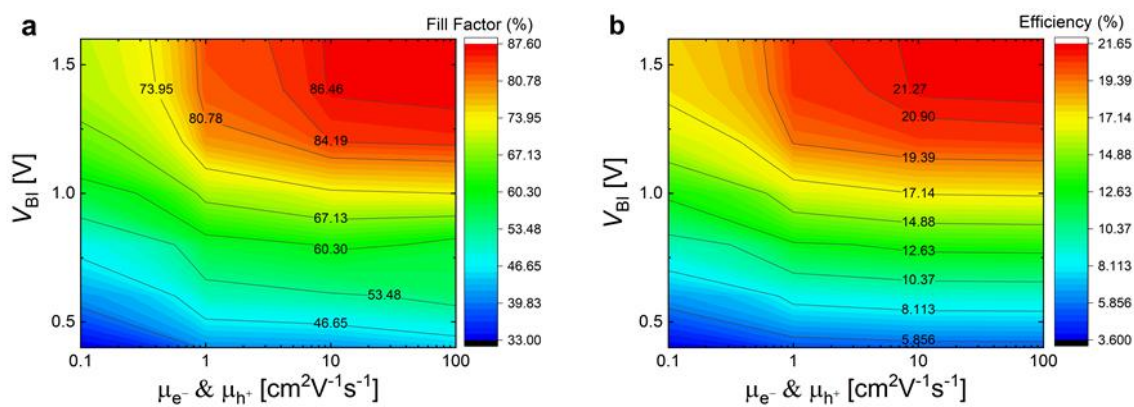

**Supplementary Figure S14.** Heat maps of fill factor (a) and PCE (b) vs. built-in voltage and electron and hole mobility in the perovskite layer ($\mu_{e^-}$ and $\mu_{h^+}$) for the standard cell. In order to reproduce the high experimental fill factor (of ~80%), a considerable $V_{BI}$ of above 1.1 V is required regardless of the perovskite mobility because it is the interface recombination that deteriorates the FF in case of low driving voltages.

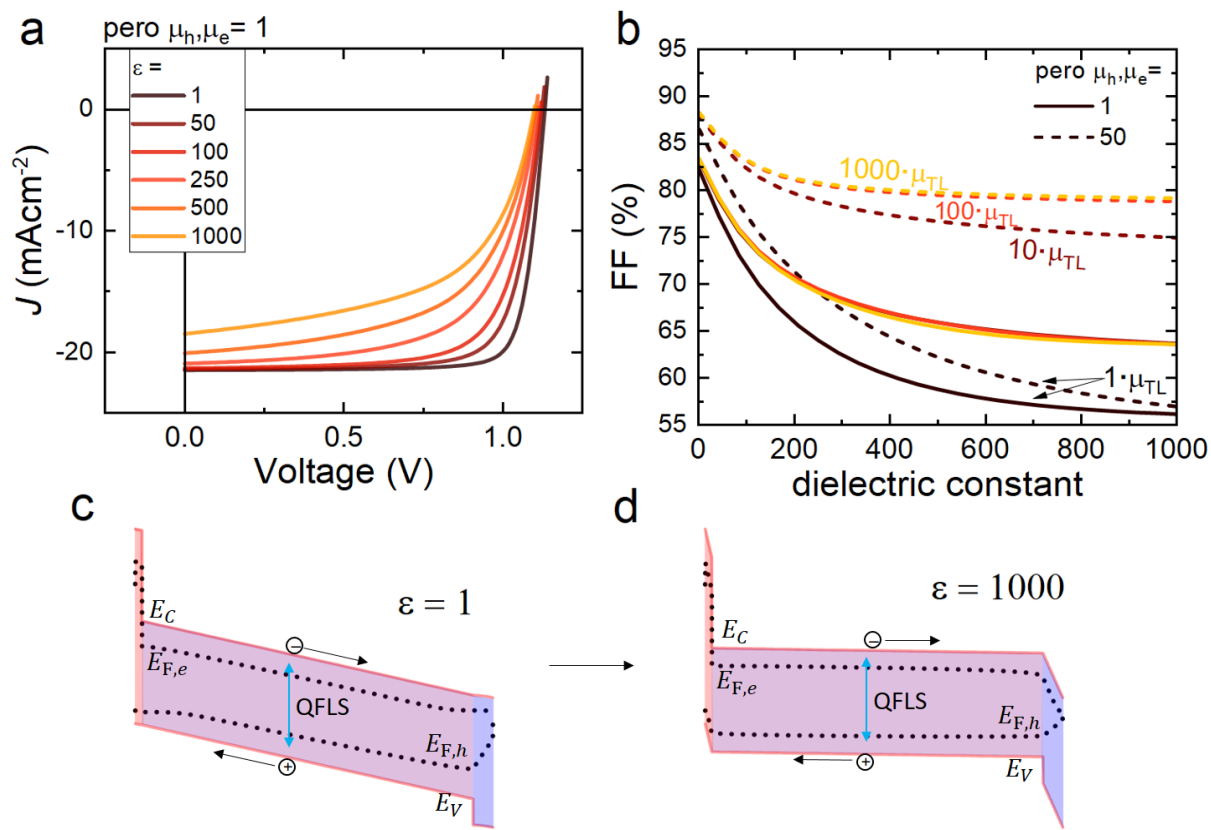

**Supplementary Figure S15.** Numerically simulated JV-curves **(a)** and fill factors (FFs) **(b)** as a function of the dielectric constant ($\epsilon_r$). Increasing $\epsilon_r$ flattens the $V_{BI}$ in the perovskite layer as shown **(c)** and **(d)**. Panel (b) shows a continuous drop of FF with increasing dielectric constant in the perovskite layer regardless of mobility in the perovskite. Note, the different coloured lines represent simulations where the mobility in both transport layers has been multiplied (increased) by the given factor. Even if the perovskite electron and hole mobility is as high as 50 cm$^2$/V/s (or even higher), a nearly flat $V_{BI}$ in the perovskite layer (here implemented by using a hypothetical $\epsilon_r$ of 1000) would not allow to reproduce the high fill factors obtained experimentally (almost 80%) if the TL mobilities are unchanged (black lines). This is a result of the increased interface recombination in this scenario. We note that this does not exclude that there are filed free regions in the perovskite in case of ions. This is because any barrier for minority carriers within the perovskite (as shown for example in Figure 2a) would significantly reduce the interface recombination which would allow a field-free perovskite layer for the given interface recombination parameters.

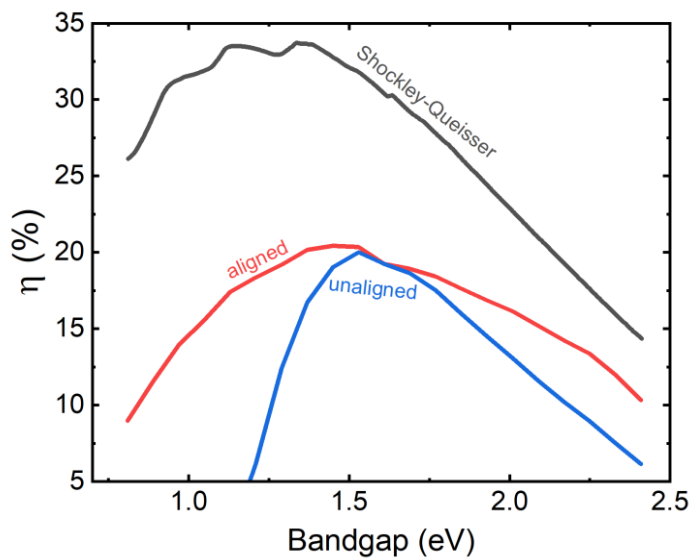

**Supplementary Figure S16.** The graphs shows a drop in PCE if the energy levels of the TLs and the work functions of the contact metals are kept as used in the standard simulations ("unaligned").

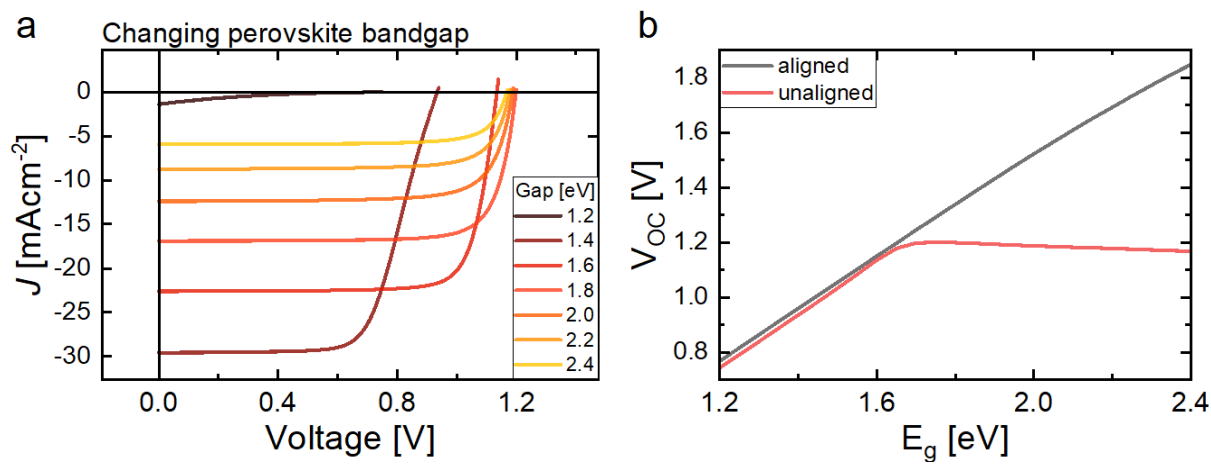

**Supplementary Figure S17. (a)** Simulated JV characteristics for different perovskite bandgaps by keeping the electron and hole transport layer energy levels constant. **(b)** The open-circuit voltage as a function of perovskite bandgap in case of energetically aligned and energetically misaligned transport layer energy levels (i.e. keeping the energy levels constant upon varying the perovskite bandgap). The graph shows that alignment of the energy level of the transport layers with respect to the perovskite is an important consideration. If only the bandgap of the perovskite is increased but the transport layer energetics is not changed than the $V_{OC}$ cannot be improved.

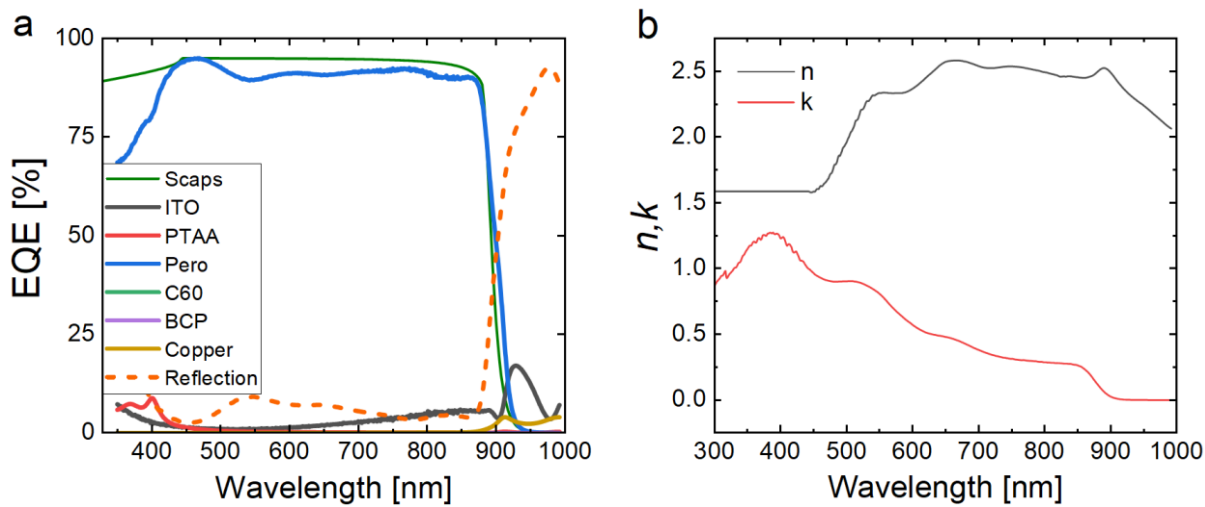

**Supplementary Figure S18.** (a) Transfer Matrix simulations of (hypothetical) optimized cells with a bandgap of 1.4 eV. Limitations on the $J_{SC}$ are only imposed by the reflections from the glass and the parasitic absorption in ITO and PTAA while the parasitic absorption of the thin PTAA layer and other layers are minimal (less than 1% average EQE loss in PTAA). However there is a relatively large reflection loss (dashed line) due to the comparatively high n of the perovskite layer[23] and the glass which can be, however, minimized by using an antireflection coating (~4% gain). Leaving only reflection from the perovskite, we obtain an EQE (solid blue line) that is on average slightly below 95% above the gap which leads to $J_{SC}$ loss of roughly ~1 mAcm$^{-2}$ as compared to the $J_{SC}$ obtained with SCAPS in the record cell (marked with an open circle in **Figure 5a**) at a bandgap of 1.4 eV. (b) shows the n and k values of the perovskite layer used for the simulations shown in panel (a). We note, the n and k values were measured on a 83-17 triple cation perovskite with a bandgap of 1.63 eV using spectroscopy ellipsometry. For the transfer matrix simulations of cells with a lower bandgap cells (1.4 eV), we used the same n and k values as for the 1.63 eV triple cation perovskite with a bandgap but shifted the x-axis accordingly to consider the available experimental data to some extend

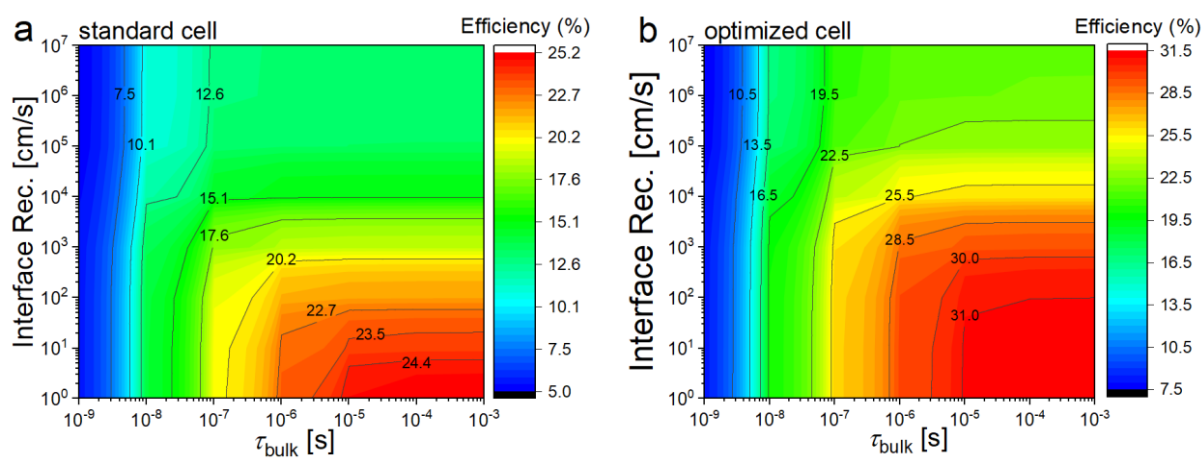

**Supplementary Figure S19.** (a) Heat maps of PCE vs. interface recombination velocity (S) (at both perovskite/transport layer interfaces) and bulk carrier lifetime ($\tau_{bulk}$) for the standard

cell **(a)** and for the high efficiency system with a perovskite bandgap of 1.4 eV and doped TLs **(b)**.

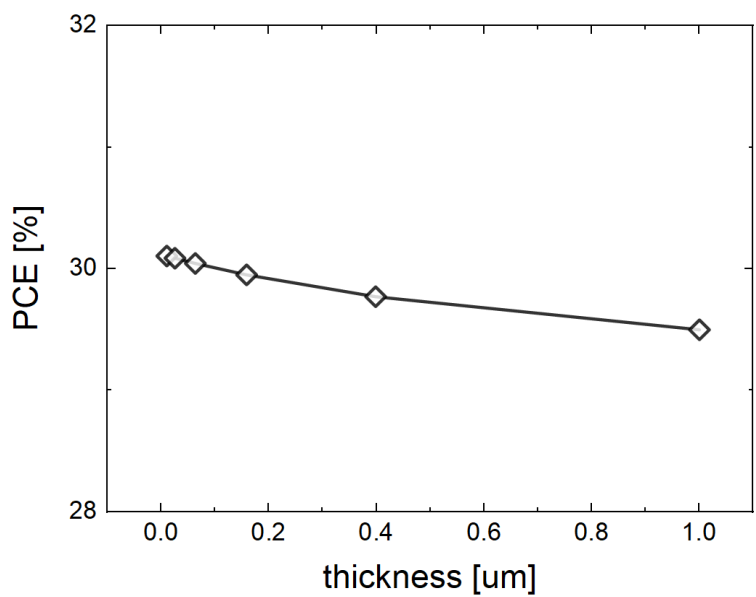

**Supplementary Figure S20.** The power conversion efficiency vs. the thickness of both transport layers with doping concentrations of $10^{19}$ cm$^{-3}$. The thickness of the TLs can be increased up a few hundred nm without significant PCE losses.

**Supplementary References**